\documentclass[times,12pt]{article}
\pdfoutput=1
\usepackage[margin=1in]{geometry}
\usepackage{amsmath,amssymb,amsfonts,latexsym,amsthm,enumerate,url}
\usepackage{booktabs,subcaption,dcolumn}
\usepackage{graphicx}
\usepackage{mathrsfs}
\usepackage{braket}
\usepackage{xcolor}
\usepackage{float}
\usepackage{hyperref}
\usepackage{mathtools}
\usepackage{adjustbox}
\usepackage{graphicx}
\date{}

\newcommand{\beq}[1]{\begin{equation}\label{#1}}
\newcommand{\enq}[0]{\end{equation}}

\newcommand{\remove}[1]{}

\newcommand{\comment}[1]{}

\title{Further Statistical Study of NISQ Experiments\footnote {Research supported by ERC grant 834735.}}
\author{Gil Kalai,  
Tomer Shoham, and Carsten Voelkmann}

\begin{document}
\maketitle

\begin{abstract}

We revisit and extend some topics that we studied in our previous works \cite {KRS22d,KRS23,KRS24} regarding the Google 2019 ``quantum supremacy" experiment. 
We extend our analysis of the prediction based on Google's digital error model (Formula (77) from Section 3 of \cite {KRS23}), based on more detailed data provided by Google.
We also provide some preliminary analysis for a few other NISQ experiments.

\end {abstract}

\section {Introduction}
\label {s:asymm}
In this paper we further study Google's 2019 quantum supremacy claim \cite {Aru+19} and especially concerns regarding the implausible predictive power of ``Formula (77)" raised in \cite {KRS23,Kal24}.
We also describe a preliminary analysis of other recent NISQ experiments. 

A central ingredient of the claims of quantum supremacy in \cite {Aru+19} was a prediction (Formula (77)) of the fidelity of samples produced by the quantum circuit based on fidelities (or probabilities of errors) for the individual components of the computer.
The small gap of 10\%--15\% reported between the prediction and the actual results appeared to be statistically surprising. 
One difficulty in studying this matter was that the Google team had not provided the individual fidelities on which their calculations were based. 
Here we further study this matter based on partial data provided by the Google team and based on retrieving the values of 2-gate fidelities from Figure 2b in \cite {Aru+19}.
Our additional analysis in Section \ref {s:77} further supports the concern, and, in particular, we could not confirm the values reported in \cite {Aru+19} and the deviations between these values and our computations are substantial.  
Appendix \ref {s:appA}  contains the reported and retrieved data on readout and gate errors in the Google experiment, Appendix \ref {s:77t} contains statistical analysis of Formula (77) based on this data. In our view, 
the assertions of \cite {Aru+19}, even those referring to the 10--20 qubit range, cannot be considered conclusive. 

In Section \ref{s:other} we discuss other NISQ experiments.
One experiment \cite {DHL+25}, by Quantinuum in cooperation with JPMorganChase, is based on trapped ions \cite{Blu+23},
another experiment \cite{Blu+23}, by researchers from Harvard/QuEra/MIT, is based on neutral atoms, and a third experiment \cite {Gao+25}, by researchers from USTC, is based on an 83-qubit random circuit.
We also briefly discuss the proposed application by Quantinuum and JPMorganChase \cite{LSN+25} to ``certified randomness."  
Appendix \ref {s:quantinuum} contains preliminary analysis of the data from Quantinuum's paper \cite {DHL+25},
Appendix \ref {s:neutral} lists data requested and 
data provided for the Harvard/QuEra/MIT neutral atom experiment \cite {Blu+23}, Appendix \ref {s:USTC_83_qubits} contains preliminary analysis of the data from the USTC 83-qubit experiment, and Appendix \ref {s:ratio} studies the ratios of 
1's in samples coming from several NISQ experiments. 
Appendix \ref {s:newexp}  lists some new NISQ experiments 
and Appendix \ref {s:wl} lists measures that would support a more accurate evaluation of experimental quantum computing.

\section {Revisiting Formula (77) for predicting fidelity}
\label {s:77}

\subsection {Formula (77) and its importance}
The Google quantum supremacy claim is based on an a priori prediction of the fidelity of a circuit based on the probabilities of error of the individual components.
Formula (77) in the supplement to \cite {Aru+19}) reads

\begin {equation}
\label {e:77}
~\widehat \phi~=~ \prod_{g \in {\cal G}_1} (1-e_g) \prod_{g \in {\cal G}_2} (1-e_g) \prod_{q \in {\cal Q}} (1-e_q).
\end {equation}

Here ${\cal G}_1$ is the set of 1-gates, ${\cal G}_2$ is the set of 2-gates, and ${\cal Q}$ is the set of qubits. 
For a gate $g$, the term $e_g$ in the formula refers to the probability of an error 
of the individual gate $g$. 
For a qubit $q$, $e_q$ is the probability of a readout error when we measure the qubit $q$.

The average value 
of 1-gate errors from Google's 2019 experiment is 
$0.0016$, of 2-gate errors is 
$0.0062$, and of readout errors is $0.038$ measured when components operate simultaneously
\cite {Aru+19}. An approximation to Formula (\ref {e:77}) based on these averaged fidelities is given by 
\begin {equation}
\label {e:77s}
{\hat \phi}^{*}= (1-0.0016)^{|{\cal G}_1|} (1-0.0062)^{|{\cal G}_2|} (1-0.038)^n .
\end {equation}

In our correspondence, the Google team proposed the following different approximation that does not take into account the 1-gates separately, but adds their contributions to a combined fidelity of the 2-gates (referred to as ``2-gate cycles"):
\begin {equation}
\label {e:77s2}
{\hat \phi}^{**}=  (1-0.0093)^{|{\cal G}_2|} (1-0.038)^n .
\end {equation}

Table 4 in \cite {KRS23} compared the prediction from Formula (77) (as reported by Google), the simplified predictions described above, and the average fidelity for the experimental circuits.

\subsection {The concern about Formula (77)} 

One concern raised in \cite {KRS23, Kal24} is that the success of the prediction of Formula (77) in the supplement of \cite {Aru+19} is statistically surprising, and this finding suggests an optimization process that invalidates the statistical claims of the article.
The paper \cite {KRS23} also provides evidence that the calibration process of the Google experiment reported in \cite {Aru+19} represents a global form of optimization toward the expected outcomes.   

In our correspondence (Nov. 2019), the Google team justified the remarkable predictive power of their Formula (77) (Equation (\ref {e:77})) with a statistical computation that is based on the following three ingredients.

\begin {enumerate}
\item
Estimations for individual readout and gate error probabilities are accurate.
The Google team reported that the error rate of their estimations of individual qubit and gate error probabilities is 20\%.
\item
Mistakes in the estimations of individual error probabilities are unbiased; namely, there are no systematic mistakes in these estimations.
\item
Readout and gate errors are statistically independent.
\end {enumerate}

Based on these assumptions, Google's (rough) estimation of the relative deviation of the prediction of Formula (77) was 
\begin {equation}
\label {e:fn}
DEV = 0.2 \cdot ( \sqrt n\cdot 0.038 + \sqrt{|{\cal G}_1|}\cdot 0.0016 + \sqrt{|{\cal G}_2|}\cdot 0.0062 ). 
\end {equation}
For example,  for $n=53$ and $m=14$, the number of 1-gates $|{\cal G}_1|=795$ and the number of 2-gates $|{\cal G}_2|=301$, and Formula \ref {e:fn} yields roughly 8.6\%. 
 
The explanation for this estimation is that once we accept that the errors in assessing the actual fidelities are unbiased (or nonsystematic), their cumulative effect grows like the square root of the number of components (as in simple random walks).
For example, for qubit errors, the probability of an error in a single qubit is, on average, $0.038(1\pm 0.2)$.
The cumulative effect of the $(1\pm 0.2)$ terms for $n$ qubits is $\sqrt n$ multiplied by 20\% of the average error, yielding $0.2 \cdot \sqrt n \cdot 0.038$. 

In \cite {KRS23} we argued that the second and third items in this explanation are implausible.
As we shall see, examining the data of the experiment indicates that the first item is incorrect as well, and the accuracy of individual readout errors appears to be considerably higher than 20\%.

\subsection {Retrieving the individual fidelities}

Google published (2020) the individual readout and 1-gate errors of their experiment (one 1-gate error is missing) in file \texttt{som\_params\_by\_qubit.csv}.
One difficulty in studying the prediction given by Formula (77) was that the Google team had not provided the individual 2-gate fidelities but only their average value. 
We asked for these values already in 2019 and the Google team promised on a few occasions to make an attempt to provide this information, but in the end they did not provide it.
Figure 2b in the Google paper \cite {Aru+19} indicates the 1-gate and 2-gate errors and based on it we were able to retrieve the 1-gate and 2-gate error rates.
For the 1-gates, in all but one case we got the same values as in the supplementary material.

{\bf Remarks:} Google's predictions based on Formula (77) were published in the files \texttt{fidelity\_4a.csv} and \texttt{fidelity\_4b.csv} 
in the data.
The Google team reported (private communication, June 2020) that their published prediction took into account small drifts in the individual values.  

The Google team uploaded additional raw data for readout errors (file \texttt{readout\_raw\_data.tar} from January 2021).
It is based on initializing the quantum computer (with many computational basis states) and then measuring.
For example, for $n=12$, there were 108,000 bitstrings generated from 36 distinct computational basis states,
and for $n=53$, there were 477,000 samples from 159 distinct states.
The average readout errors based on these data are shown in parentheses in Table \ref{t:google-readout-data-av}.

\subsection {Analysis based on individual fidelities}

We can summarize our findings as follows. 

\begin {enumerate}
\item 
We were not able to confirm the values of Formula (77)'s predictions reported in the Google paper.
When we use the individual error rates reported by the Google paper (including those retrieved from Figure 2b of \cite {Aru+19}) we do not get the predicted values reported in the Google paper.
(This is the case for both versions of the readout errors provided by the Google team.)
The deviations from the reported values from Formula (77) are substantial.

\item
There are substantial gaps between the two files with readout values (provided by the Google team in 2020 and 2021, respectively).
The source of the gaps is not clear and they cast doubt on the claim that all these values are stable up to 20\%.
For the 2021 data, the gaps between readout errors for the same qubits are substantial for data generated for different numbers of qubits. 

\item
Using the individual readout errors gave us an opportunity to examine the effect of replacing the average value of readout errors with the individual readout errors.
As it turns out, the average value itself changed as a function of the number of qubits. 
In Table \ref{t:readourerr-77} we give the effect of the readout errors on fidelity when the average readout error rate (0.038) is replaced by the individual readout errors in the detailed Google file \texttt{som\_params\_by\_qubit.csv} from January 2020.
Overall, this change takes us away both from the reported (77) predictions (which we are unable to check) and from the average ${\cal F}_{\mathrm{XEB}}$ value. (See Table \ref {t:adj-77}.)
\end {enumerate}

\subsection {The concern regarding the patch circuits}

The Google 2019 experiment included patch circuits that consisted of two circuits on two sets of qubits (patches) without any 2-gates acting on one qubit in one patch and one qubit in the other.
For these circuits, the XEB fidelity estimator was computed separately for the two patches and then multiplied.
For the patch circuits of type {\bf EFGH}, the XEB fidelities were very close to the XEB fidelities of the full circuits (of the same size) and were systematically $\sim$10\% lower than the values predicted by Formula (77). (This is noted in \cite {Aru+19} where it is attributed to the calibration process.) We observed in \cite {KRS23} (Section 4.6) that for the first patch the XEB values were close to the XEB value of full circuits of the same size while for the second patch it was lower by 10\%-15\%. (See Table \ref {tab:merged_table} in Appendix \ref {s:77t}.) The individual fidelities provided by the Google data (and Figure 2b in \cite {Aru+19}) do not give an explanation for this finding. For patch circuits of type {\bf ABCDCDAB} ($n=53$, depth 12--20), the XEB fidelities (namely, the product of the XEB values for the two patches) were, again, very close to the Formula (77) prediction for the {\it full circuits}: the deviations were $\sim \pm$ 5\%--10\% (see Table 5 in \cite {KRS23}). 
However, for the individual patches themselves there were large deviations (in some cases $\sim\pm$ 40\%--50\%) between the XEB values and the predictions by Formula (77). (See Table \ref {tab:merged_table} and Figure \ref {fig:patch_vs_full} in Appendix \ref {s:77t}.)

\section {Other NISQ experiments}
\label {s:other}

\subsection {Quantinuum's random quantum circuit experiments}

Quantinuum's 2024 paper \cite {DHL+25} provides random circuit sampling data taken with their H2 trapped-ion quantum processor.
They employ random geometry (RG) circuits, i.e., arbitrary/flexible 2-gate connectivity as opposed to 2D nearest-neighbor geometry for Google’s Sycamore and Willow processors. 

In this experiment (as in earlier experiments on quantum volume), the sample sizes are small (between 20 to 100 bitstrings per circuit compared to 500K in Google's 2019 experiment) and samples are drawn from many different circuits.
(One reason for the small samples is that due to physical ion transport, the processor needs 80~ms (milliseconds) per 2-gate layer (with $n=56$), as opposed to 20~ns (nanoseconds) for Google’s superconducting circuits.)
Additional information on the Quantinuum experiment and preliminary statistical analysis of the data are given in Appendix \ref {s:quantinuum}.

Quantinuum's methodology is to supply very short samples for hundreds of circuits.
(Having lower error rates and hence higher fidelity compared to superconducting quantum computers enable shorter samples.)
Having very small samples does not allow testing the empirical distribution, and working with many different circuits makes methodologically flawed optimization toward the desired outcomes easier and harder to be detected compared to the Google experiment.
This concern is especially relevant for the application to certified randomness.

Two quick remarks: First, Quantinuum's paper \cite {DHL+25} asserts that unlike previous RCS demonstrations, no special purpose compilation or calibration was used. Second, in 2024 we asked the Quantinuum team for the availability of larger samples but were told that the sample size was optimal for their purposes.

\subsubsection *{Certified randomness} Certified randomness is based on the assumption that the source of the random bits cannot be trusted.
The concern is that the source is adversarial and would feed you the output of a pseudorandom generator instead of truly random bits.
(If the source of the random bits can be trusted, there are easy methods to extract randomness from physical processes.)
When certified randomness is based on quantum supremacy that in turn is based on interpolations from small circuits to large ones, this procedure cannot be trusted for two reasons:
first, the extrapolation argument is heuristic and the quality of circuits may deteriorate beyond expectation with their size.
Second, and more important, the results for smaller circuits may reflect classical simulation rather than genuine quantum circuits.
This means that for an adversarial agent who tries to present ``certified" random bits based on quantum supremacy, it is relatively easy to present fake quantum supremacy claims, and, at present, we do not have methods to certify the validity of the quantum supremacy claims themselves.\footnote {It is plausible that ``certification" for sampling problems such as random circuit sampling is computationally intractable not only for classical computers but also for quantum computers.}

\subsection {The Harvard/QuEra neutral atom experiments: data requests}
A recent 2024 experiment \cite {Blu+23} by a group from Harvard, MIT, and QuEra, 
drew considerable attention.
The experiment describes logical circuits based on neutral atoms and is considered among the most important breakthroughs in experimental quantum computing in recent years.
In Appendix \ref {s:neutral}, we describe our requests for data on this experiment and list the data that were provided by the authors.
(See \cite {KRS22d} for a similar description of our requests and the authors' responses for the Google 2019 experiment.)

Some partial data was provided by the authors in February 2024 and preliminary statistical findings based on it were described in \cite {KRS24}.
In particular, we noticed strong stability of the Fourier--Walsh transform, not observed in other data from experiments and simulations, which we found surprising for empirical data. 
The data on the various quantum error correction codes that we requested were not provided.
We were told that it would take a long time to compile and package the data, that the researchers are currently working on new experiments, and that they will be packaging everything ahead of time for their future papers.
For further details, 
see Appendix \ref {s:neutral}.

\subsection {USTC's 83-qubit random circuit sampling}

Gao et al. \cite {Gao+25} from USTC describe random circuit sampling experiments with 31 and 83 superconducting qubits on their Zuchongzhi 3.0 quantum processor.
Their full (i.e., 1-patch), 2-patch, and 4-patch circuits have depths between 12 and 32.

In order to model their measured XEB fidelities, they use a refined version of Google's Formula~(77) discrete-error model.
Their refined model includes not only factors for the individual 1-gate, 2-gate, and readout fidelities, but also the following two factors: 
(1) qubit-specific fidelities of the idle (i.e., identity) 1-gates of those qubits that are not part of a 2-gate during a 2-gate layer
and (2) qubit-specific fidelities corresponding to an imperfect preparation of the initial state $\lvert 000\ldots\rangle$ (called ``P0 error” in \cite {Gao+25}).

In contrast to Google's 2019 ``quantum supremacy" paper, Formula~(77) overestimates USTC's measured XEB fidelities by a factor between 1.2 and 3.7.
The consideration of the additional two factors of the refined discrete-error model leads to a strong agreement with USTC’s measured XEB fidelity for the 31-qubit circuits, for which the ratio ``discrete-error model / XEB" is between 0.8 and 1.
For the 83-qubit circuits, this ratio is between 0.5 and 0.7.
One reason for this deviation might lie in the factor for the idle gates, which was calculated using the average (and not the qubit-specific) value of the errors, since ~\cite {Gao+25} does not include this qubit-specific information.

Table~\ref{t:USTC_discrete_error_model} in Appendix \ref {s:USTC_83_qubits} compares USTC's measured XEB fidelities with Google's Formula~(77) predictions and with USTC's refined discrete-error model predictions.

\subsection {Ratios of 1's in NISQ samples}

In Appendix \ref {s:ratio} we study the ratios of 1's across several NISQ experiments \cite{Aru+19,Mor+24,Gao+25}. We note that in some cases the ratio of 1's increases with depth, and for some experiments it is above 0.5. 

\section {Conclusion} 

Our further study of Google's Formula (77) strengthens the concern raised in \cite {KRS23}  about the methodology of the optimization process that led to the remarkable agreement between the predicted values of the XEB fidelities and the statistical estimations based on the empirical samples.
First, we have not been able to reproduce (or even come close to reproducing) the values obtained from Formula (77) reported in the Google paper.
Second, using the data provided by Google, {\it reduced} the quality of the prediction relative to rougher estimates based on average values.
Third, when we combine Google’s data with the two-gate error rates extracted from Figure 2b of the Google paper, the resulting predictions do not match the numbers reported by Google---and they also do not match the empirical XEB fidelities.
In addition, the readout data supplied by Google is inconsistent with the claim made by Google’s researchers that the error rates are stable to within $\pm$20\%, and it is also inconsistent with the assumption that the error rates do not themselves exhibit systematic errors. The error rates supplied by Google, together with those retrieved from Figure 2b of \cite {Aru+19}, do not explain the behavior of the XEB fidelities for patched circuits.

Beyond revisiting Google’s Formula (77), we have also reported on preliminary examinations of  additional quantum-computing experiments \cite {Blu+23,DHL+25,Gao+25,Mor+24}. 

Overall, the findings reported in this paper underscore the importance of conducting detailed experiments and providing large samples already in the 10--20 qubit range, as well as the importance of supplying full and accurate data about the experiments. In our view, at present, the assertions of \cite {Aru+19}, even those referring to the 10--20 qubit range, cannot be considered conclusive.

\subsection *{Acknowledgements}
Research supported by ERC grant 834735. 
We thank Yosi Rinott for helpful discussions.


\clearpage
\appendix
\begin{center}
{\LARGE\bfseries Supplementary Materials}
\end{center}

\vspace{1ex} 

\section{Google 2019 experiment: Readout and gate errors}
\label {s:appA}
\enlargethispage{4cm}
\subsection{Readout error rate reported by Google}

\begin{table}[H]
\begin{subtable}[t]{0.49\textwidth}
\scalebox{0.85}{%
\begin{tabular}{||c c c c||} 
\hline
qubit & $n_{\text{ins}}$ & $q_{0\rightarrow 1}$ & $q_{1 \rightarrow 0}$\\ [0.5ex] 
\hline \hline
q0\_5&38&0.50 (1.19)&2.78 (8.48)\\
q0\_6&40&0.51 (2.40)&3.07 (5.23)\\
q1\_4&36&2.97 (1.28)&3.52 (4.73)\\
q1\_5&28&0.81 (1.31)&3.01 (4.25)\\
q1\_6&28&1.13 (2.28)&6.55 (5.48)\\
q1\_7&30&0.76 (1.72)&3.63 (5.65)\\
q2\_4&14&1.40 (1.92)&2.47 (4.81)\\
q2\_5&14&3.34 (2.58)&2.82 (6.52)\\
q2\_6&18&2.03 (3.10)&5.81 (6.18)\\
q2\_7&24&0.98 (1.02)&3.74 (4.43)\\
q2\_8&34&0.49 (5.11)&4.06 (5.99)\\
q3\_2&36&1.97 (2.73)&5.60 (5.13)\\
q3\_3&12&3.82 (2.33)&5.07 (7.15)\\
q3\_4&12&2.30 (0.86)&6.08 (5.37)\\
q3\_5&12&0.74 (2.09)&7.47 (6.88)\\
q3\_6&12&0.36 (3.37)&4.16 (5.40)\\
q3\_7&20&1.28 (1.28)&3.93 (3.72)\\
q3\_8&34&1.03 (0.61)&3.77 (4.87)\\
q3\_9&46&0.88 (7.47)&5.08 (11.21)\\
q4\_1&38&1.93 (4.71)&5.07 (7.02)\\
q4\_2&22&1.05 (1.84)&3.58 (4.79)\\
q4\_3&12&1.86 (5.53)&5.90 (5.20)\\
q4\_4&12&3.24 (1.27)&6.25 (7.85)\\
q4\_5&12&5.29 (1.93)&15.57 (11.21)\\
q4\_6&12&0.44 (1.00)&10.52 (5.03)\\
q4\_7&20&1.46 (1.48)&7.80 (4.32)\\
q4\_8&44&0.93 (0.46)&5.99 (7.65)\\
\hline
\end{tabular}}
    \end{subtable}   
\begin{subtable}[t]{0.49\textwidth}
\scalebox{0.85}{%
\begin{tabular}{||c c c c||} 
\hline
qubit & $n_{\text{ins}}$ & $q_{0\rightarrow 1}$&$q_{1 \rightarrow 0}$\\ [0.5ex] 
\hline \hline
q4\_9&50&1.45 (1.16)&6.06 (4.18)\\
q5\_0&53&1.29 (1.48)&5.61 (9.43)\\
q5\_1&32&1.89 (1.24)&5.42 (5.01)\\
q5\_2&22&1.16 (6.62)&9.37 (6.61)\\
q5\_3&12&2.44 (1.14)&5.86 (4.25)\\
q5\_4&12&1.83 (1.85)&5.08 (4.11)\\
q5\_5&12&4.95 (2.01)&6.77 (6.25)\\
q5\_6&12&12.96 (0.44)&7.46 (4.45)\\
q5\_7&42&1.39 (1.00)&4.31 (3.28)\\
q5\_8&51&1.34 (2.22)&4.56 (4.02)\\
q6\_1&32&2.87 (1.60)&7.01 (2.86)\\
q6\_2&24&3.08 (1.25)&3.27 (5.71)\\
q6\_3&18&1.14 (1.25)&5.63 (5.28)\\
q6\_4&16&3.89 (4.30)&4.33 (5.18)\\
q6\_5&16&1.26 (1.30)&3.67 (4.35)\\
q6\_6&39&9.16 (0.85)&4.57 (3.96)\\
q6\_7&48&1.82 (1.39)&3.74 (4.61)\\
q7\_2&30&1.75 (0.82)&4.20 (3.90)\\
q7\_3&26&2.99 (6.07)&4.13 (4.76)\\
q7\_4&26&3.04 (1.00)&4.77 (3.01)\\
q7\_5&43&1.60 (0.50)&6.22 (4.14)\\
q7\_6&47&3.08 (1.35)&9.24 (4.51)\\
q8\_3&41&3.34 (1.44)&7.45 (5.65)\\
q8\_4&45&2.04 (3.46)&5.86 (7.77)\\
q8\_5&49&2.69 (6.24)&7.14 (7.67)\\
q9\_4&53&1.50 (8.02)&4.88 (13.05)\\
\textbf{Avg}&&\textbf{2.25 (2.32)}&\textbf{5.47 (5.71)}\\
\textbf{Std}&&\textbf{2.12 (1.88)}&\textbf{2.25 (2.07)}\\
\hline    
\end{tabular}}
\end{subtable}
\caption{\small {Sycamore's asymmetric readout error rates in \% per qubit under simultaneous operation
(source: Google's file \texttt{som\_params\_by\_qubit.csv} in agreement with the lower two panels of Figure S24 in the supplement of \cite {Aru+19}).
The values in parentheses are the relative frequencies of the asymmetric readout errors from Google's initialization/measurement experiments for $n=53$
(source: Google's file \texttt{readout\_raw\_data.tar}).
The column $n_{\text{ins}}$ is the number $n$ of qubits, for which the respective qubit is added to the circuit
(source: Python list QUBIT\_ORDER in the circuit description files of the full circuits).
}}
\label{t:google-readout-data}
\end{table}

\newpage
\subsubsection*{Average readout error rates for Google's Sycamore}

\begin{table}[H]
\begin{center} \resizebox{0.85\textwidth}{!}{
    \begin{tabular}{||c | c c c c | c||} 
 \hline
  $n$&\multicolumn{4}{|c|}{Average readout error in \%}&Empirical \\
&$q_{0\rightarrow 1}$&$q_{1 \rightarrow 0}$ & Avg & Diff & Diff \\ [0.5ex] 
\hline \hline
12&3.35 (2.12)&7.18 (5.04)&5.27 (3.58)&3.83 (2.92)&2.51\\
14&3.21 (2.17)&6.53 (4.60)&4.87 (3.39)&3.32 (2.43)&2.26\\
16&3.13 (2.08)&6.22 (4.55)&4.68 (3.32)&3.09 (2.47)&2.24\\
18&2.96 (2.04)&6.16 (4.97)&4.56 (3.51)&3.20 (2.93)&2.45\\
20&2.80 (2.27)&6.13 (4.59)&4.47 (3.43)&3.33 (2.32)&2.06\\
22&2.65 (2.16)&6.16 (5.00)&4.41 (3.58)&3.51 (2.84)&2.51\\
24&2.60 (2.19)&5.94 (5.14)&4.27 (3.67)&3.34 (2.95)&2.68\\
26&2.63 (2.34)&5.83 (5.24)&4.23 (3.79)&3.20 (2.90)&2.51\\
28&2.51 (2.05)&5.75 (5.17)&4.13 (3.61)&3.24 (3.12)&2.57\\
30&2.43 (2.20)&5.63 (5.15)&4.03 (3.68)&3.20 (2.95)&2.57\\
32&2.42 (2.20)&5.67 (5.07)&4.05 (3.64)&3.25 (2.87)&2.58\\
34&2.33 (2.32)&5.56 (5.74)&3.95 (4.03)&3.23 (3.42)&2.41\\
36&2.33 (2.20)&5.51 (5.15)&3.92 (3.68)&3.18 (2.95)&2.66\\
38&2.27 (2.18)&5.42 (5.40)&3.85 (3.79)&3.15 (3.22)&2.89\\
39&2.45 (2.16)&5.40 (5.53)&3.93 (3.85)&2.95 (3.37)&2.78\\
40&2.40 (2.16)&5.34 (5.66)&3.87 (3.91)&2.94 (3.50)&3.03\\
41&2.43 (2.27)&5.40 (5.51)&3.92 (3.89)&2.97 (3.24)&2.98\\
42&2.40 (2.36)&5.37 (5.47)&3.89 (3.92)&2.97 (3.11)&2.68\\
43&2.38 (2.35)&5.39 (5.50)&3.89 (3.93)&3.01 (3.15)&3.17\\
44&2.35 (2.26)&5.40 (5.53)&3.88 (3.90)&3.05 (3.27)&2.89\\
45&2.34 (2.40)&5.41 (5.59)&3.88 (4.00)&3.07 (3.19)&3.00\\
46&2.31 (2.43)&5.41 (5.58)&3.86 (4.01)&3.10 (3.15)&2.80\\
47&2.33 (2.56)&5.49 (5.75)&3.91 (4.16)&3.16 (3.19)&2.87\\
48&2.32 (2.74)&5.45 (5.76)&3.89 (4.25)&3.13 (3.02)&2.83\\
49&2.32 (2.63)&5.49 (5.80)&3.91 (4.22)&3.17 (3.17)&2.91\\
50&2.31 (2.73)&5.50 (5.58)&3.91 (4.16)&3.19 (2.85)&2.74\\
51&2.29 (2.61)&5.48 (5.55)&3.89 (4.08)&3.19 (2.94)&2.81\\
53&2.25 (2.33)&5.47 (5.70)&3.86 (4.02)&3.22 (3.37)&2.93\\
\hline
\end{tabular}}
\caption{Columns $q_{0\rightarrow1}$ and $q_{1\rightarrow0}$ are the average readout error rates based on Table~\ref {t:google-readout-data}, each average includes only the qubits in this circuit.
In parenthesis are the readout error relative frequencies in Google's initialization/measurement file \texttt{readout\_raw\_data.tar} uploaded by Google in January 2021. (For every value of $n$ we took the file corresponding to this value.)  
The rightmost column gives the difference between percentage of zeroes and ones in the Google experimental samples, all full circuits. 
This difference is smaller than the difference based on the readout data.
We do not witness in the Sycamore experimental data additional asymmetry predicted in \cite {FGG+23} that is based on non-unital gates.}
\label{t:google-readout-data-av}
\end{center}
\end{table}

Further remarks about Table \ref {t:google-readout-data-av}:
The raw data files provided by the Google team are based on initializing Sycamore (with many computational basis states) and then measuring the qubits.
For example, for $n=12$, there were 108,000 bitstrings generated from 36 distinct computational basis states,
and for $n=53$, there were 477,000 samples from 159 distinct states.
We assume that the readout errors provided in the supplementary data are based on a similar method.
We do not know the reason for the differences between the readout error parameters from the supplementary data (file \texttt{som\_params\_by\_qubit.csv}) and from the additional raw data (file \texttt{readout\_raw\_data.tar}).   

\subsection{Reported and retrieved gate errors}
\label {s:gate_errors}

We calculated the individual 2-gate errors from their colors in Figure~2b of \cite{Aru+19}.
For this we read out the RGB (red, green, blue) color values for each tick mark (0.0008, 0.0009, ..., 0.01) and the top (0.017) of the color legend of Figure~2b in \cite {Aru+19}.
We then interpolated 99 RGB values between each pair of neighboring tick marks, e.g., between the error values 0.0008 and 0.0009.
This resulted in interpolated RGB values for 100 error values in the interval $[0.0008, 0.0009[$, for another 100 error values in $[0.0009, 0.001[$, and so on.

\begin{figure*}[!ht]
    \centering
    \includegraphics[width=4.5in]{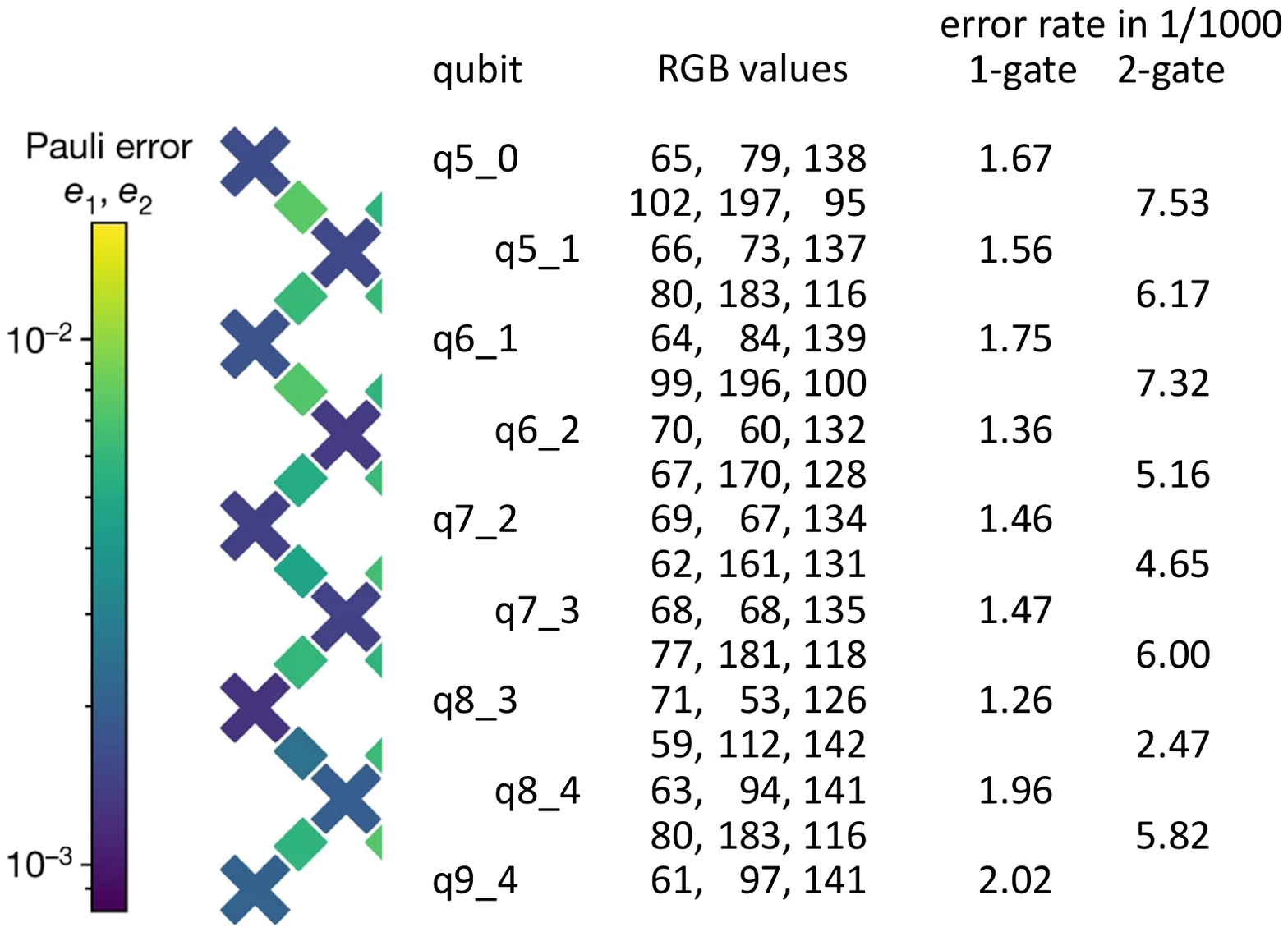}
    \caption{RGB values and gate error rates for some exemplary Sycamore-53 qubits and gates retrieved from Figure~2b in \cite {Aru+19}.} 
    \label{fig_RGB}
\end{figure*}

There are 12 such tick mark intervals overall, with 100 error values in each interval.
The procedure thus assigned RGB values to 1201 error values in the entire error range $[0.0008, 0.017]$ of the color legend.
We read out the RGB values for every 1-gate and 2-gate in Figure 2b of \cite {Aru+19} and assigned to each gate the unique one of the 1201 error values whose L1-distance in 3-dimensional RGB space to the gate's color is minimal.
Examples of the retrieved RGB values and resulting error rates are shown in Figure~\ref{fig_RGB}.

\begin{table}[H]
\centering
\setlength{\tabcolsep}{2pt}
\renewcommand{\arraystretch}{0.95}
\begin{adjustbox}{max width=5.7in}
\begin{tabular}{||c c||@{\hspace{10pt}}c c||@{\hspace{10pt}}c c||@{\hspace{10pt}}c c||}
\hline
\textbf{qubit} & \textbf{1-gate error} &
\textbf{qubit} & \textbf{1-gate error} &
\textbf{qubit} & \textbf{1-gate error} &
\textbf{qubit} & \textbf{1-gate error} \\
\hline\hline
q0\_5 & 2.5 & q3\_5 & 1.2 & q4\_9 & 1.4 & q6\_4 & 2.0 \\
q0\_6 & 1.3 & q3\_6 & 1.5 & q5\_0 & 1.6 & q6\_5 & 1.3 \\
q1\_4 & 1.5 & q3\_7 & 1.3 & q5\_1 & 1.6 & q6\_6 & 1.4 \\
q1\_5 & 1.6 & q3\_8 & 2.1 & q5\_2 & 1.5 & q6\_7 & 1.0 \\
q1\_6 & 3.2 & q3\_9 & 3.4 & q5\_3 & 1.4 & q7\_2 & 1.4 \\
q1\_7 & 1.2 & q4\_1 & 1.7 & q5\_4 & 1.8 & q7\_3 & 1.5 \\
q2\_4 & 1.0 & q4\_2 & 1.1 & q5\_5 & 1.3 & q7\_4 & 2.0 \\
q2\_5 & 1.0 & q4\_3 & 1.4 & q5\_6 & 1.1 & q7\_5 & 1.8 \\
q2\_6 & 1.8 & q4\_4 & 1.7 & q5\_7 & 1.1 & q7\_6 & 2.7 \\
q2\_7 & 1.3 & q4\_5 & 0.8 & q5\_8 & 2.5 & q8\_3 & 1.3 \\
q2\_8 & 1.3 & q4\_6 & 0.9 & q6\_1 & 1.8 & q8\_4 & 2.0 \\
q3\_2 & 1.7 & q4\_7 & 1.3 & q6\_2 & 1.4 & q8\_5 & 1.6 \\
q3\_3 & 1.3 & q4\_8 & 2.3 & q6\_3 & 1.2 & q9\_4 & 1.6 (2.0) \\
q3\_4 & 1.2 &        &     &        &     &        &      \\
\hline
\end{tabular}
\end{adjustbox}
\caption{\textbf{Sycamore-53: 1-gate error rates in 1/1000 (simultaneous operation)}
arranged in four blocks (each with columns \emph{qubit} and \emph{1-gate error}).
Aggregates: Average = \textbf{1.58}, Median = \textbf{1.4}, Min = \textbf{0.8}, Max = \textbf{3.4}.
Except for qubit q9\_4, the 1-gate errors are taken from Google's file \texttt{som\_params\_by\_qubit.csv}, column \texttt{sq\_errors\_simultaneous},
which are identical to the values in the Supplement to \cite {Aru+19}, Figure~S23, center right panel ``XEB error (simultaneous)", and which are consistent with the colors in Figure~2b of \cite {Aru+19}.
For qubit q9\_4, the 1-gate errors reported by Google are inconsistent: this value is missing in the file \texttt{som\_params\_by\_qubit.csv}, it is stated as $1.6 \times 1/1000$ in Figure~S23, center right panel in the Supplement to \cite {Aru+19}, and its color in Figure~2b of \cite {Aru+19} yields a value of $2.0 \times 1/1000$.
Since Google does not provide the individual 2-gate errors, we calculated them from their colors in Figure~2b of \cite {Aru+19}.}
\label{table:error_rates_1gate}
\end{table}

\begin{table}[t]
\centering
\setlength{\tabcolsep}{2pt}
\renewcommand{\arraystretch}{0.85}
\begin{adjustbox}{max width=5.6in}
\begin{tabular}{||c c c||@{\hspace{12pt}}c c c||@{\hspace{12pt}}c c c||}
\hline
\textbf{qubit1} & \textbf{qubit2} & \textbf{2-gate error} &
\textbf{qubit1} & \textbf{qubit2} & \textbf{2-gate error} &
\textbf{qubit1} & \textbf{qubit2} & \textbf{2-gate error} \\
\hline\hline
q0\_5 & q0\_6 & 9.54  & q3\_7 & q3\_8 & 5.32  & q5\_4 & q6\_4 & 7.88 \\
q0\_5 & q1\_5 & 3.03  & q3\_7 & q4\_7 & 4.58  & q5\_5 & q5\_6 & 5.00 \\
q0\_6 & q1\_6 & 8.85  & q3\_8 & q3\_9 & 3.80  & q5\_5 & q6\_5 & 7.88 \\
q1\_4 & q1\_5 & 3.21  & q3\_8 & q4\_8 & 6.29  & q5\_6 & q5\_7 & 5.91 \\
q1\_4 & q2\_4 & 2.37  & q3\_9 & q4\_9 & 7.00  & q5\_6 & q6\_6 & 9.19 \\
q1\_5 & q1\_6 & 9.54  & q4\_1 & q4\_2 & 4.88  & q5\_7 & q5\_8 & 3.42 \\
q1\_5 & q2\_5 & 2.81  & q4\_1 & q5\_1 & 5.62  & q5\_7 & q6\_7 & 3.80 \\
q1\_6 & q1\_7 & 2.88  & q4\_2 & q4\_3 & 4.31  & q6\_1 & q6\_2 & 7.32 \\
q1\_6 & q2\_6 & 6.00  & q4\_2 & q5\_2 & 4.31  & q6\_2 & q6\_3 & 6.24 \\
q1\_7 & q2\_7 & 8.42  & q4\_3 & q4\_4 & 6.61  & q6\_2 & q7\_2 & 5.16 \\
q2\_4 & q2\_5 & 4.52  & q4\_3 & q5\_3 & 6.06  & q6\_3 & q6\_4 & 8.94 \\
q2\_4 & q3\_4 & 8.00  & q4\_4 & q4\_5 & 5.77  & q6\_3 & q7\_3 & 6.61 \\
q2\_5 & q2\_6 & 7.00  & q4\_4 & q5\_4 & 8.25  & q6\_4 & q6\_5 & 9.54 \\
q2\_5 & q3\_5 & 6.24  & q4\_5 & q4\_6 & 7.21  & q6\_4 & q7\_4 & 16.91 \\
q2\_6 & q2\_7 & 7.32  & q4\_5 & q5\_5 & 4.76  & q6\_5 & q6\_6 & 6.61 \\
q2\_6 & q3\_6 & 5.16  & q4\_6 & q4\_7 & 10.00 & q6\_5 & q7\_5 & 6.71 \\
q2\_7 & q2\_8 & 5.06  & q4\_6 & q5\_6 & 6.17  & q6\_6 & q6\_7 & 7.06 \\
q2\_7 & q3\_7 & 4.15  & q4\_7 & q4\_8 & 3.65  & q6\_6 & q7\_6 & 4.36 \\
q2\_8 & q3\_8 & 4.53  & q4\_7 & q5\_7 & 4.20  & q7\_2 & q7\_3 & 4.65 \\
q3\_2 & q3\_3 & 5.16  & q4\_8 & q4\_9 & 15.05 & q7\_3 & q7\_4 & 5.82 \\
q3\_2 & q4\_2 & 6.11  & q4\_8 & q5\_8 & 2.81  & q7\_3 & q8\_3 & 6.00 \\
q3\_3 & q3\_4 & 4.47  & q5\_0 & q5\_1 & 7.53  & q7\_4 & q7\_5 & 9.00 \\
q3\_3 & q4\_3 & 6.43  & q5\_1 & q5\_2 & 6.17  & q7\_4 & q8\_4 & 6.24 \\
q3\_4 & q3\_5 & 5.23  & q5\_1 & q6\_1 & 6.17  & q7\_5 & q7\_6 & 9.69 \\
q3\_4 & q4\_4 & 5.91  & q5\_2 & q5\_3 & 6.00  & q7\_5 & q8\_5 & 9.19 \\
q3\_5 & q3\_6 & 9.39  & q5\_2 & q6\_2 & 5.82  & q8\_3 & q8\_4 & 2.47 \\
q3\_5 & q4\_5 & 4.53  & q5\_3 & q5\_4 & 5.00  & q8\_4 & q8\_5 & 7.38 \\
q3\_6 & q3\_7 & 5.62  & q5\_3 & q6\_3 & 6.61  & q8\_4 & q9\_4 & 5.82 \\
q3\_6 & q4\_6 & 3.65  & q5\_4 & q5\_5 & 6.24  &      &      &      \\
\hline
\end{tabular}
\end{adjustbox}
\caption{\textbf{Sycamore-53: 2-gate error rates in 1/1000 (simultaneous operation)}
arranged in three blocks (each with columns \emph{qubit1}, \emph{qubit2}, \emph{2-gate error}).
Aggregates: Average = \textbf{6.23}, Median = \textbf{6.00}, Min = \textbf{2.37}, Max = \textbf{16.91}.
Since Google does not provide the individual 2-gate errors, we calculated them from their colors in Figure~2b of \cite{Aru+19}.
For this we read out the RGB (red, green, blue) color values for each tick mark and the top of the color legend of Figure~2b in \cite {Aru+19}.
We then interpolated 99 RGB values between each pair of neighboring tick marks, resulting in 1201 values overall.
We read out the RGB values for every 1-gate and 2-gate in Figure 2b of \cite {Aru+19} and assigned to each gate the unique one of the 1201 error values whose L1-distance in 3-dimensional RGB space to the gate's color is minimal.
A more detailed description of this procedure is given at the beginning of Section~\ref{s:gate_errors}.}
\label{table:error_rates_2gate}
\end{table}

\section{Revisiting Formula (77): tables}
\label {s:77t}
\enlargethispage{4cm}


\begin{table}[H]
\begin{center}
\begin{adjustbox}{width=0.65\textwidth}
\begin{tabular}{||c c c c||} 
\hline
    & reported qubit-specific & qubit-specific     & \\ [0.5ex]
$n$ & readout errors          & readout err. freq. & $(1-0.038)^n$\\ [0.5ex] 
\hline\hline
12&0.520&0.645&0.628\\
14&0.495&0.620&0.581\\
16&0.462&0.583&0.538\\
18&0.429&0.529&0.498\\
20&0.399&0.492&0.461\\
22&0.369&0.447&0.426\\
24&0.349&0.407&0.395\\
26&0.323&0.366&0.365\\
28&0.305&0.357&0.338\\
30&0.289&0.321&0.313\\
32&0.265&0.303&0.289\\
34&0.253&0.243&0.268\\
36&0.235&0.259&0.248\\
38&0.223&0.229&0.229\\
39&0.208&0.215&0.221\\
40&0.204&0.201&0.212\\
41&0.193&0.193&0.204\\
42&0.188&0.186&0.196\\
43&0.180&0.177&0.189\\
44&0.174&0.173&0.182\\
45&0.167&0.157&0.175\\
46&0.162&0.153&0.168\\
47&0.152&0.136&0.162\\
48&0.148&0.122&0.156\\
49&0.141&0.120&0.150\\
50&0.136&0.118&0.144\\
51&0.132&0.117&0.139\\
53&0.123&0.113&0.128\\
\hline
\end{tabular}
\end{adjustbox}
\end{center}
\caption{Probability of no readout error for Sycamore with different number of qubits $n$.
The second column is based on the qubit-specific readout error rates in Google's file \texttt{som\_params\_by\_qubit.csv}
(for each qubit, the average of the two asymmetric readout error rates with simultaneous operation).
The third column is based on the qubit-specific readout error relative frequencies of initialization/measurement experiments in Google's file \texttt{readout\_raw\_data.tar}
(for every value of $n$ we took the file corresponding to this value, and for each qubit, the average of its two asymmetric readout error relative frequencies was used).
The rightmost column is based on the reported average readout error rate 0.038 (for all 53 qubits).}
\label{t:readourerr-77}
\end{table}

\begin{table}[H]
\begin{center}
\hspace*{-0.5cm}
\begin{adjustbox}{width=16.5cm}
\begin{tabular}{||c|r c c c c c c |c c|c c c|c c||} 
\hline
& \emph{Eq.} & (\ref {e:77s}) & (\ref {e:77}) & (\ref {e:77}) & (\ref {e:77}) & (\ref {e:77}) & (\ref {e:77}) & (\ref {e:77}) & (\ref {e:77}) & (\ref {e:77s2}) & (\ref {e:77s2}) & (\ref {e:77s2}) &  & \\[0.5ex]
$n$ & \emph{1-gate}  & avg & qb-spec & avg & qb-spec & avg & qb-spec & qb-spec & qb-spec & – & – & – & (77)  & ${\cal F}_{\mathrm{XEB}}$ \\
    & \emph{2-gate}  & avg & avg & avg & avg & avg & avg & gate-spec & gate-spec & – & – & – &  & \\
& \emph{readout} & avg & avg & qb-spec & qb-spec & rel.freq. & rel.freq. & qb-spec & rel.freq. & avg & qb-spec & rel.freq. &  & \\[0.5ex]
\hline\hline
12&&0.3242&0.3423&0.2685&0.2835&0.3329&0.3514&0.2878&0.3568&0.3586&0.2970&0.3682&0.3862&0.3694\\
14&&0.2687&0.2888&0.2286&0.2457&0.2866&0.3080&0.2497&0.3131&0.3023&0.2571&0.3224&0.3320&0.3297\\
16&&0.2228&0.2390&0.1915&0.2055&0.2412&0.2589&0.2040&0.2569&0.2548&0.2190&0.2759&0.2828&0.2720\\
18&&0.1801&0.1938&0.1553&0.1671&0.1913&0.2060&0.1639&0.2019&0.2069&0.1784&0.2198&0.2207&0.2442\\
20&&0.1483&0.1611&0.1284&0.1394&0.1585&0.1722&0.1356&0.1674&0.1728&0.1495&0.1846&0.1875&0.2184\\
22&&0.1222&0.1339&0.1057&0.1159&0.1280&0.1403&0.1143&0.1384&0.1443&0.1248&0.1512&0.1554&0.1650\\
24&&0.0988&0.1091&0.0873&0.0964&0.1019&0.1125&0.0954&0.1114&0.1171&0.1036&0.1208&0.1256&0.1407\\
26&&0.0819&0.0900&0.0725&0.0797&0.0820&0.0901&0.0763&0.0863&0.0987&0.0874&0.0989&0.1024&0.1141\\
28&&0.0679&0.0728&0.0612&0.0657&0.0716&0.0769&0.0628&0.0735&0.0832&0.0751&0.0878&0.0907&0.0949\\
30&&0.0549&0.0594&0.0508&0.0550&0.0563&0.0610&0.0534&0.0592&0.0676&0.0625&0.0693&0.0759&0.0823\\
32&&0.0452&0.0488&0.0414&0.0447&0.0473&0.0510&0.0432&0.0494&0.0564&0.0517&0.0590&0.0624&0.0713\\
34&&0.0372&0.0401&0.0351&0.0378&0.0338&0.0364&0.0370&0.0356&0.0471&0.0445&0.0428&0.0459&0.0586\\
36&&0.0301&0.0324&0.0286&0.0308&0.0314&0.0338&0.0310&0.0340&0.0383&0.0363&0.0399&0.0454&0.0520\\
38&&0.0249&0.0265&0.0243&0.0258&0.0249&0.0264&0.0264&0.0271&0.0322&0.0314&0.0322&0.0369&0.0419\\
\hline
39&&0.0224&0.0239&0.0211&0.0225&0.0219&0.0233&0.0228&0.0237&0.0291&0.0274&0.0283&0.0307&0.0328\\
40&&0.0202&0.0215&0.0194&0.0207&0.0191&0.0204&0.0206&0.0203&0.0262&0.0252&0.0248&0.0275&\\
41&&0.0186&0.0200&0.0176&0.0189&0.0176&0.0189&0.0187&0.0187&0.0245&0.0232&0.0232&0.0251&\\
42&&0.0167&0.0181&0.0160&0.0173&0.0158&0.0171&0.0173&0.0171&0.0221&0.0211&0.0209&0.0229&0.0265\\
43&&0.0150&0.0162&0.0144&0.0155&0.0141&0.0152&0.0153&0.0150&0.0199&0.0190&0.0186&0.0204&0.0191\\
44&&0.0135&0.0144&0.0129&0.0138&0.0129&0.0137&0.0138&0.0138&0.0179&0.0172&0.0171&0.0190&\\
45&&0.0122&0.0129&0.0116&0.0123&0.0109&0.0116&0.0125&0.0117&0.0161&0.0154&0.0145&0.0160&\\
46&&0.0111&0.0115&0.0107&0.0111&0.0101&0.0104&0.0113&0.0107&0.0150&0.0144&0.0136&0.0149&\\
47&&0.0100&0.0102&0.0094&0.0096&0.0084&0.0085&0.0097&0.0087&0.0135&0.0127&0.0113&0.0128&\\
48&&0.0090&0.0092&0.0086&0.0088&0.0071&0.0072&0.0089&0.0073&0.0122&0.0116&0.0095&0.0113&\\
49&&0.0081&0.0083&0.0076&0.0078&0.0065&0.0067&0.0078&0.0067&0.0109&0.0103&0.0088&0.0109&\\
50&&0.0073&0.0075&0.0068&0.0070&0.0060&0.0061&0.0068&0.0060&0.0099&0.0093&0.0081&0.0093&\\
51&&0.0065&0.0066&0.0062&0.0063&0.0055&0.0056&0.0062&0.0055&0.0089&0.0084&0.0075&0.0090&\\
\hline
53&&0.0055&0.0056&0.0053&0.0053&0.0048&0.0049&0.0053&0.0048&0.0077&0.0074&0.0068&0.0077&0.0074\\
\hline
\end{tabular}
\end{adjustbox}
\end{center}
\caption  {\small {A priori predictions for the fidelity based on Google's formula (77) for different number of qubits $n$.
The term ``avg" in the various columns refers to the average over all 53 qubits or over all 86 qubit pairs. 
The columns under ``1-gate: avg" contain the fidelities calculated with Google's reported average single-qubit gate error (simultaneous) 0.16\%.
The columns under ``1-gate: qb-spec" are based on qubit-specific single-qubit gate errors reported in Google's file \texttt{som\_params\_by\_qubit.csv}, column \texttt{sq\_errors\_simultaneous}.
The columns under ``2-gate: avg" contain the fidelities calculated with Google's reported average 2-qubit gate error (simultaneous) 0.62\%.
The columns under ``2-gate: gate-spec" are based on gate-specific 2-qubit gate errors extracted from their colors in Figure~2b of \cite {Aru+19}.
The columns under ``readout: avg" contain the fidelities calculated with Google's reported average readout error (simultaneous) 3.8\%.
The columns under ``readout: qb-spec" are based on qubit-specific readout errors reported in Google's file \texttt{som\_params\_by\_qubit.csv}, columns \texttt{sq\_readout\_error\_\textbar0\textrangle{}} and \texttt{sq\_readout\_error\_\textbar1\textrangle}.
The columns under ``readout: rel.freq." are based on qubit-specific readout error relative frequencies in initialization/measurement experiments in Google's 
file \texttt{readout\_raw\_data.tar}.
Columns \eqref {e:77s2} are similarly based on Equation \eqref {e:77s2}.
The column (77) shows Google's prediction based on Formula (77).
The rightmost column shows the average XEB fidelity.}}
\label{t:adj-77}
\end{table}

\subsection *{Data for patch circuits} 

\begin{table}[H]
\centering
\resizebox{\textwidth}{!}{%
\begin{tabular}{||c c c c c c c c c c c c c||} 
\hline
 \multicolumn{3}{|c|}{Circuit} & \multicolumn{3}{c|}{First Patch} & \multicolumn{3}{c|}{Second Patch} & \multicolumn{2}{c|}{Combined circuit} & \multicolumn{2}{c|}{Google} \\ \hline \hline 
n & m & Type & n & (77) & XEB  & n & (77) & XEB & (77) & XEB & Pred & Adj. pred \\
\hline\hline
12&14& EFGH &6&0.5832&0.5771&6&0.5416&0.5547&0.3159&0.3201&0.3862&0.43195\\
14&14& EFGH &6&0.5832&0.5677&8&0.4700&0.5759&0.2741&0.3269&0.3320&0.37133\\
16&14& EFGH &8&0.4764&0.4928&8&0.4700&0.5506&0.2239&0.2713&0.2828&0.31630\\
18&14& EFGH &9&0.4275&0.4707&9&0.4208&0.5323&0.1799&0.2506&0.2207&0.24684\\
20&14& EFGH &9&0.4275&0.4722&11&0.3482&0.4798&0.1489&0.2266&0.1875&0.20971\\
22&14& EFGH &11&0.3604&0.3825&11&0.3482&0.4635&0.1255&0.1773&0.1554&0.17381\\
24&14& EFGH &12&0.3275&0.3293&12&0.3198&0.4213&0.1047&0.1387&0.1256&0.14048\\
26&14& EFGH &14&0.2620&0.2670&12&0.3198&0.4279&0.0838&0.1142&0.1024&0.11453\\
28&14& EFGH &14&0.2620&0.2695&14&0.2630&0.3509&0.0689&0.0946&0.09072&0.10147\\
30&14& EFGH &15&0.2406&0.2470&15&0.2435&0.3270&0.0586&0.0808&0.07594&0.08494\\
32&14& EFGH &17&0.1947&0.2131&15&0.2435&0.3326&0.0474&0.0709&0.06236&0.06975\\
34&14& EFGH &17&0.1947&0.2068&17&0.2088&0.2437&0.0407&0.0504&0.04592&0.05136\\
36&14& EFGH &18&0.1756&0.1898&18&0.1936&0.2632&0.0340&0.0500&0.04535&0.05072\\
38&14& EFGH &19&0.1593&0.1611&19&0.1818&0.2426&0.0290&0.0391&0.03693&0.04130\\
39&14& EFGH &20&0.1414&0.1432&19&0.1818&0.2436&0.0257&0.0349&0.03073&0.03502\\
40&14& EFGH &20&0.1414&0.1394&20&0.1641&0.2197&0.0232&0.0306&0.02751&0.03135\\
41&14& EFGH &21&0.1289&0.1044&20&0.1641&0.2177&0.0212&0.0227&0.02514&0.02865\\
42&14& EFGH &21&0.1289&0.1149&21&0.1512&0.1930&0.0195&0.0222&0.0229&0.02609\\
43&14& EFGH &22&0.1139&0.0972&21&0.1512&0.1841&0.0172&0.0179&0.02043&0.02328\\
44&14& EFGH &22&0.1139&0.0955&22&0.1363&0.164&0.01550&0.0157&0.01901&0.02166\\
45&14& EFGH &23&0.1032&0.0879&22&0.1363&0.1643&0.0141&0.0144&0.01600&0.01823\\
46&14& EFGH &23&0.1032&0.1004&23&0.1238&0.1509&0.0128&0.0152&0.01485&0.01692\\
47&14& EFGH &24&0.0883&0.0876&23&0.1238&0.1493&0.0109&0.0131&0.01281&0.01460\\
48&14& EFGH &24&0.0883&0.0931&24&0.1172&0.1347&0.0103&0.0125&0.01134&0.01325\\
49&14& EFGH &25&0.0774&0.0688&24&0.1172&0.1448&0.0091&0.0100&0.01088&0.01271\\
50&14& EFGH &25&0.0774&0.0723&25&0.1018&0.1264&0.0079&0.0091&0.00934&0.01091\\
51&14& EFGH &25&0.0774&0.0717&26&0.0931&0.1170&0.0072&0.0084&0.00898&0.01049\\
\hline
53&12&ABCD&27&0.0748&0.0821&26&0.1054&0.1563&0.0079&0.0128&0.01206&0.01374\\
53&14&ABCD&27&0.0657&0.0647&26&0.0931&0.1325&0.0061&0.0086&0.00793&0.00926\\
53&16&ABCD&27&0.0572&0.0486&26&0.0840&0.1113&0.0048&0.0054&0.00531&0.00632\\
53&18&ABCD&27&0.0498&0.0354&26&0.0758&0.0855&0.0038&0.0030&0.00356&0.00432\\
53&20&ABCD&27&0.0437&0.0261&26&0.0669&0.0697&0.0029&0.0018&0.00234&0.00291\\
\hline
\end{tabular}%
}
\caption{Different estimates of the fidelity for the patch circuit of types {\bf EFGH} and {\bf ABCDCDAB}. Columns marked (77) represent computations 
based on the error rates reported by Google in 2020 and those retrieved from Figure 2b of \cite {Aru+19}. 
For the combined circuits the values are the products of those for the individual patches. 
The two rightmost columns are Google's reported prediction based on Formula (77) for the full circuit, and this prediction adjusted to the removal of the two-gates in the patch circuits, that is Adj. pred = $(1-0.0062)^{-a} \cdot$ Pred, where $a$ is the number of two-gates removed, and $0.0062$ is the average error of the two-gates.
The XEB values reported by Google largely agree with their (77) prediction for full circuits, see Table 5 in \cite {KRS23}.
(Our computations somewhat differ from Google's.)}
\label{tab:merged_table}
\end{table}

\begin{figure}[H]
	\begin{center}
\includegraphics[width=0.85\textwidth]{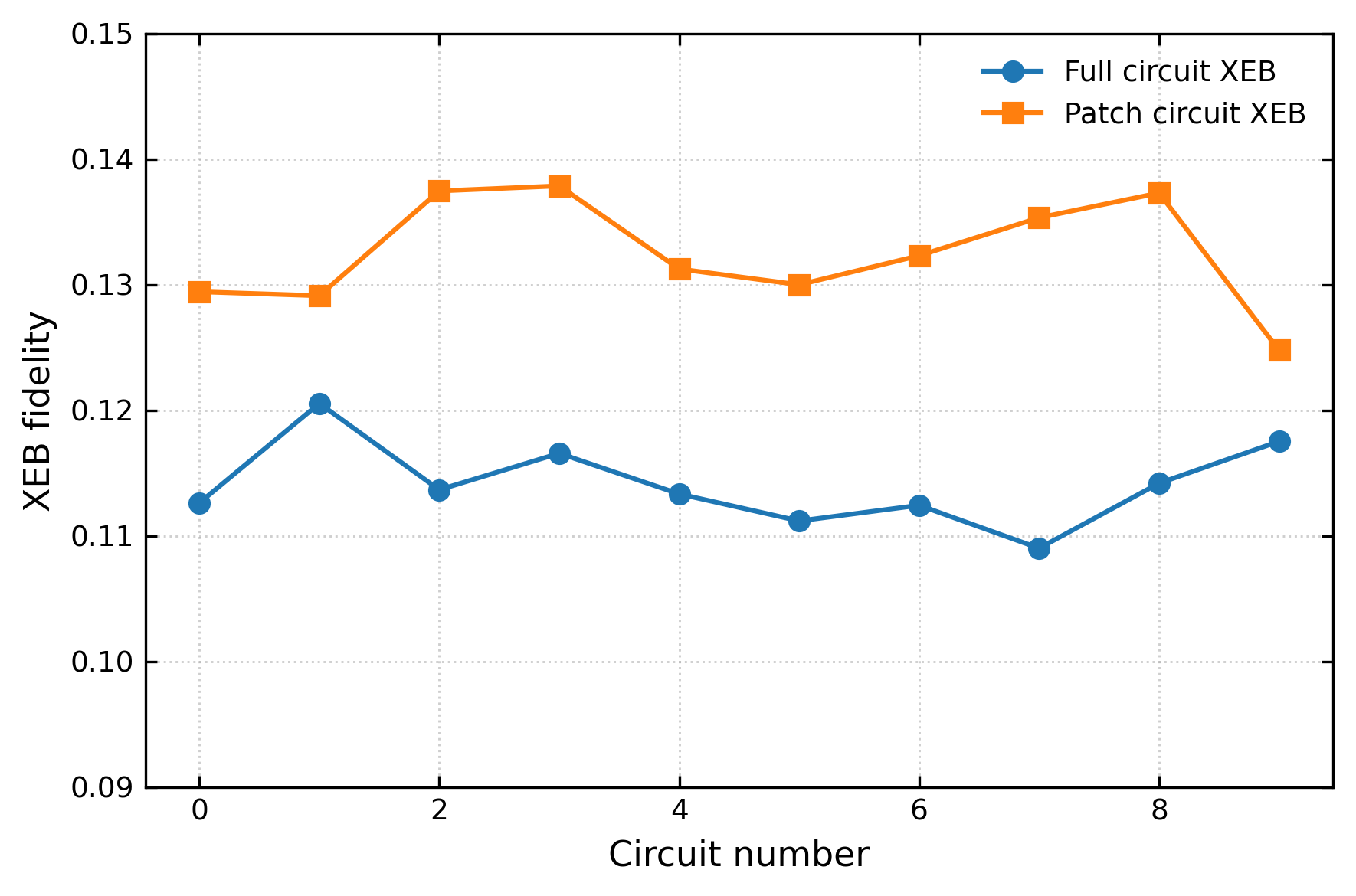}
		\caption{This figure shows the XEB values for the ten full circuits (blue dots) with 26 qubits and the ten 26-qubit patches for the ten patch circuits with 53 qubits and depth 14 (orange dots). 
        The values of the ten orange circuits are consistently higher than those of the ten blue circuits. 
        What could be the physical or engineering reason that the 26-qubit patches exhibit systematically higher XEB values?
        }
		\label{fig:patch_vs_full}
	\end{center}
\end{figure}
\bigskip
\newpage
\section{Random circuit sampling with Quantinuum's H2 trapped-ion quantum processor}
\label{s:quantinuum}
\subsection {Background}
Quantinuum's 2024 publication \cite {DHL+25} provides random circuit sampling data taken with their H2 trapped-ion quantum processor.
They employ random-geometry (RG, random-graph) circuits, i.e., arbitrary/flexible 2-gate connectivity as opposed to 2D nearest-neighbor geometry for Google’s Sycamore and Willow processors.
Due to physical ion transport, the processor needs 80~ms (milliseconds) per 2-gate layer (with $n=56$), as opposed to 20~ns (nanoseconds) for Google’s superconducting Sycamore and Willow.
This $10^6$ factor generates relatively small numbers of sampled bitstrings compared to Google and therefore to restricted statistical analysis.

Their data include 2530 circuits ($n=16-56$, depth $4-96$) with a total of 49,800 measured bitstrings, including

\begin{itemize}
\item 1000 ``XEB” circuits ($n=16-56$, depth $8-20$),\\
20 bitstrings in each circuit;\\
amplitudes for all sampled bitstrings of 550 of these circuits ($n=16-40$, depth $8-20$), no further amplitudes for other circuits
\item 1050 ``MB”-circuits (mirror-benchmarking), including
\begin{itemize}
\item 1000 circuits with 20 bitstrings in each circuit\\
($n=16-56$, depth $8-20$),
\item 50 circuits with 100 bitstrings in each circuit\\
($n=56$, depth 24)
\end{itemize}
\item 480 “Transport\_1QRB”-circuits ($n=16-56$, depth $4-96$),\\
10 bitstrings in each circuit,\\
transport 1-gate randomized benchmarking (standard Clifford gate randomization) intended to capture the combined effects of memory errors and 1-gate errors (i.e., all errors not originating from the 2Q gates), no 2-gates.
\end{itemize}

\subsubsection *{A Request from the Quantinuum team}

In June 2024 we had a brief correspondence with the Quantinuum team.
In particular, we asked: ``Can you provide larger samples for your quantum circuits? A few hundred bitstrings are too little a sample for a variety of statistical purposes."
The response was: ``All data from each test is shared in the GitHub repository. The number of circuits and shots for each test were selected to maximize the chance of passing the test with the minimum amount of runtime."
In our opinion, larger samples will be required to understand the Quantinuum quantum computer.

\subsection {Preliminary statistical finding}
\begin{figure*}[!ht]
    \centering
    \includegraphics[width=4.2in]{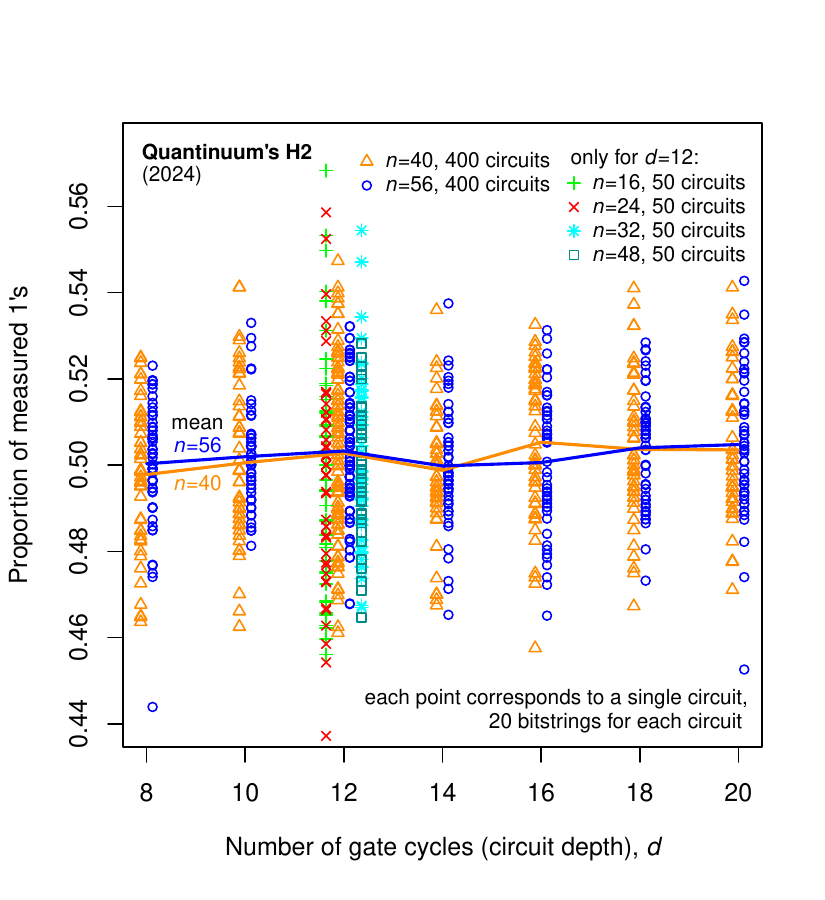}
    \caption{\textbf{Proportion of measured 1's as a function of circuit depth for Quantinuum's H2 trapped-ion quantum processor \cite {DHL+25}.}
    Each data point corresponds to one of the 1000 ``XEB” circuits.}
    \label{fig_Quantinuum_proportion_ones_vs_depth}
\end{figure*}

Figure~\ref{fig_Quantinuum_proportion_ones_vs_depth} shows that there is no statistically significant deviation from $1/2$ in the proportion of measured 1's.
This is in contrast to the superconducting architecture of Google's Sycamore-53 (where the proportion of 1's is less than $1/2$) and USTC's Zuchongzhi-56 (where the proportion of 1's is greater than $1/2$). For further details see Appendix \ref {s:ratio}.

\begin{figure*}[!ht]
    \centering
    \includegraphics[width=4.2in]{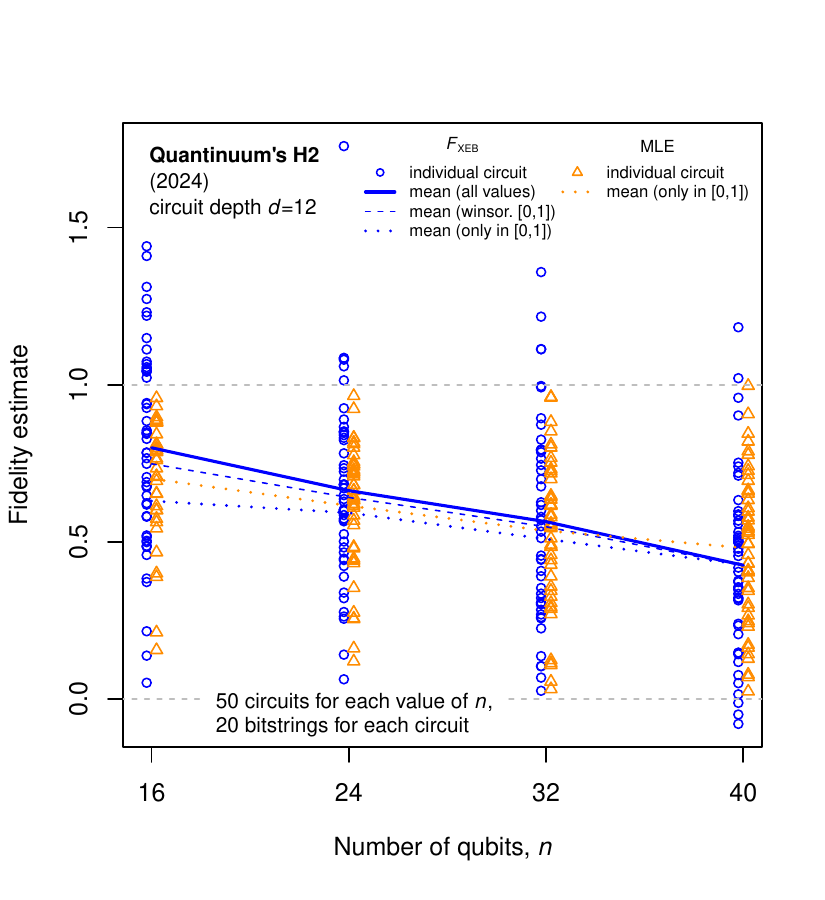}
    \caption{\textbf{XEB and MLE fidelity estimates as a function of circuit size (i.e., the number of qubits, $n$) for Quantinuum's H2 trapped-ion quantum processor \cite {DHL+25}.}
    The maximum likelihood fidelity estimate (MLE) is calculated by Eq.~(4.14) in \cite {RSK22} and is restricted to the interval $[0,1]$.
    For some of the circuits, no MLE solution in [0, 1] exists; these MLE data points are not shown.
    The different XEB mean values are calculated by using: (1) all values (both within and outside the interval [0,1]), (2) winsorized (values greater than 1 are regarded as 1, negative values are regarded as 0), (3) only the values within the interval [0,1].}
    \label{fig_Quantinuum_XEB_MLE_d12}
\end{figure*}

Figure~\ref{fig_Quantinuum_XEB_MLE_d12} recreates Quantinuum's XEB fidelity estimates of their Figure~9b in \cite {DHL+25} based on their published data.
In addition, it also includes the maximum likelihood fidelity estimate (MLE, Eq.~(4.14) in \cite {RSK22}).
Due to the comparatively small number of 20 bitstrings for each of the circuits shown, the spread of the fidelity estimate values is considerable.
The XEB fidelity estimator shows values greater than 1 and less than 0 for some of the 200 circuits shown.

\begin{figure*}[!ht]
    \centering
    \includegraphics[width=4.2in]{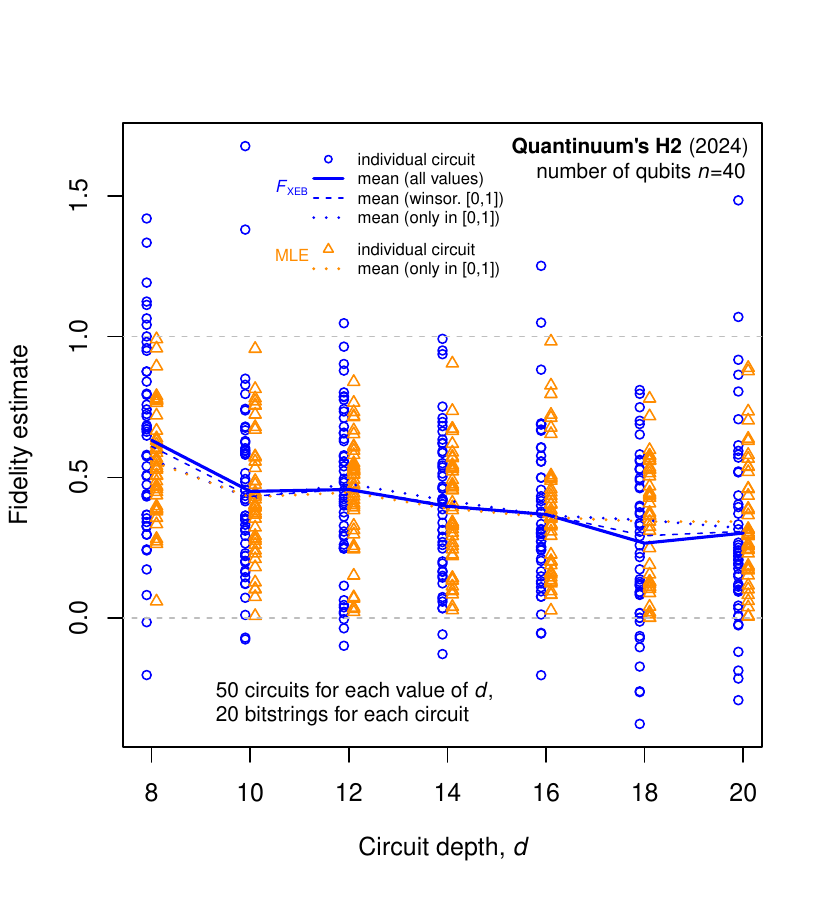}
    \caption{\textbf{XEB and MLE fidelity estimates as a function of circuit depth for Quantinuum's H2 trapped-ion quantum processor \cite {DHL+25}.}
    As in Figure~\ref{fig_Quantinuum_XEB_MLE_d12}, the MLE data points without an MLE solution in [0, 1] are not shown.}
    \label{fig_Quantinuum_XEB_MLE_N40}
\end{figure*}

Analogously, Figure~\ref{fig_Quantinuum_XEB_MLE_N40} recreates Quantinuum's Figure~9c in \cite {DHL+25} based on their published data.
Again, the MLE fidelity estimate is shown for each circuit.

\begin{figure*}[!ht]
    \centering
    \includegraphics[width=4.2in]{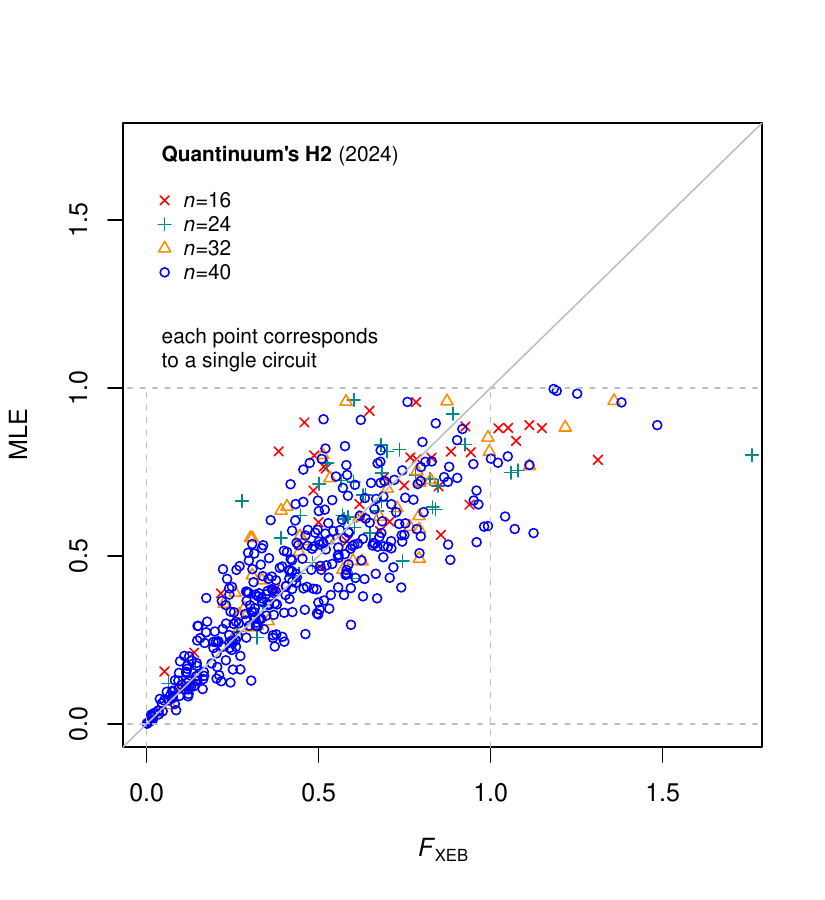}
    \caption{\textbf{Fidelity estimates MLE vs. XEB for Quantinuum's H2 trapped-ion quantum processor \cite {DHL+25}.}
    Circuits without an MLE solution in the interval [0,1] are not shown.}
    \label{fig_Quantinuum_XEB_vs_MLE}
\end{figure*}

The relation between the MLE and the XEB fidelity estimates for each of the ``XEB” circuits is reported in Figure~\ref{fig_Quantinuum_XEB_vs_MLE}.

\clearpage  

\section{The Harvard/QuEra experiment: data requests}
\label {s:neutral}
A recent 2024 experiment \cite {Blu+23} by a group from Harvard, MIT, and QuEra drew considerable attention.
The experiment describes logical circuits based on neutral atoms and is considered among the most important breakthroughs in experimental quantum computing in recent years.

In this appendix, we describe our requests for data on this experiment and list the data that were provided by the authors.
(See \cite {KRS22d} for a similar description of our requests and the authors' responses for the Google 2019 experiment.)
We requested the data represented in the figures in the article and suggested that the full data be uploaded to a server.
Specifically, we asked (mainly in February 2024) for the following data:

\begin {itemize}
\item Logical and physical samples for the IQP circuits with 12, 24, and 48 qubits, and the description of the circuits.

\item Data for the Steane code data from Figure 3 of \cite{Blu+23}.

\item The two-copy measurement data from Figure 6 of \cite{Blu+23}).

\item The data for 3- 5-, and 7- surface codes and the transversal CNOT gates based on them 
from Figure 2 of \cite{Blu+23} (and other figures as well). 

\item The data for the multi-dimensional code (Figure 5 of \cite{Blu+23}).

\item Details about error rates for gates and qubits.

\end {itemize}

In February 2024 we received the data for the 12-qubit IQP logical circuits and in \cite {KRS24} we made preliminary analysis of this data.
Later we got the logical and physical samples for the IQP circuits, with descriptions (in Qiskit) of the circuits themselves.
The data for the various quantum error correction codes were not provided and
we were told that it would take long time to compile and package the data, that the researchers were currently working on new experiments, and that, following the experience from this paper, they would be packaging everything ahead of time for their future papers.  

\section {USTC's 83-qubit random circuit sampling}
\label{s:USTC_83_qubits}
Table~\ref{t:USTC_discrete_error_model} shows modeled and measured fidelities for the circuits of USTC's 31-qubit and 83-qubit random circuit sampling experiments conducted in 2024 \cite {Gao+25}.
The modeled fidelities are calculated using Google's Formula (77) and USTC's refined discrete-error model.
The latter includes two additional factors: one for idle 1-gates of qubits not used in 2-gates and one for preparation errors.

\begin{table}[H]
\begin{center}
\hspace*{-1cm}
\begin{adjustbox}{width=18.5cm}
\begin{tabular}{||c c c|c c c|c|c c|c|c c|c c c||} 
\hline
 & number & number & 1-gate & 2-gate & readout & Formula & idle gate & prep. & modeled & measured & modeled & ratio & ratio & ratio \\[0.5ex]
circuit & of & of & fidelity & fidelity & fidelity & (77) & fidelity & fidelity & fidelity & XEB in & fidelity & (77)/ & fid.calc./ & fid.calc./ \\[0.5ex]
type & qubits & cycles & & & & & (avg.) & (avg.) & (calc.) & figures & in figures & XEB & XEB & mod. fid. \\[0.5ex]
 & $n$ & $m$ & in \% & in \% & in \% & in \% & in \% & in \% & in \% & in \% & in \% & & & in fig. \\[0.5ex]
 & & & (A) & (B) & (C) & (D) & (E) & (F) & (G) & (H) & (I) & (D/H) & (G/H) & (G/I) \\[0.5ex]
\hline\hline
full&31&12&70.0&62.4&75.5&33.0&76.7&81.6&20.6&27.3&24.7&1.21&0.76&0.83\\
full&31&16&62.7&53.3&75.5&25.2&70.2&81.6&14.5&19.4&17.3&1.30&0.75&0.84\\
full&31&20&56.2&45.6&75.5&19.3&64.3&81.6&10.1&13.2&12.1&1.47&0.77&0.84\\
full&31&24&50.3&38.9&75.5&14.8&58.9&81.6&7.1&9.5&8.5&1.56&0.75&0.84\\
full&31&28&45.1&33.3&75.5&11.3&53.9&81.6&5.0&6.5&6.0&1.73&0.76&0.84\\
full&31&32&40.4&28.4&75.5&8.7&49.3&81.6&3.5&4.6&4.2&1.87&0.75&0.84\\
\hline
full&83&12&34.9&22.5&48.5&3.8&65.1&56.1&1.4&–&–&–&–&–\\
full&83&16&25.2&13.7&48.5&1.7&56.5&56.1&0.53&–&–&–&–&–\\
full&83&20&18.3&8.3&48.5&0.74&48.9&56.1&0.20&–&–&–&–&–\\
full&83&24&13.2&5.1&48.5&0.32&42.4&56.1&0.077&–&–&–&–&–\\
full&83&28&9.5&3.1&48.5&0.14&36.8&56.1&0.030&–&–&–&–&–\\
full&83&32&6.9&1.9&48.5&0.063&31.9&56.1&0.011&–&0.023&–&–&0.49\\
\hline
2-patch&31&12&70.0&63.7&75.5&33.6&76.9&81.6&21.1&25.2&24.4&1.33&0.84&0.86\\
2-patch&31&16&62.7&54.8&75.5&25.9&70.5&81.6&14.9&18.0&17&1.44&0.83&0.88\\
2-patch&31&20&56.2&47.1&75.5&20.0&64.6&81.6&10.5&12.1&11.8&1.66&0.87&0.89\\
2-patch&31&24&50.3&40.5&75.5&15.4&59.2&81.6&7.4&8.9&8.3&1.73&0.83&0.90\\
2-patch&31&28&45.1&34.9&75.5&11.9&54.2&81.6&5.3&6.6&5.7&1.79&0.79&0.91\\
2-patch&31&32&40.4&30.0&75.5&9.2&49.7&81.6&3.7&4.6&4.0&2.00&0.81&0.93\\
\hline
2-patch&83&12&34.9&24.7&48.5&4.2&61.9&56.1&1.5&–&–&–&–&–\\
2-patch&83&16&25.2&15.5&48.5&1.9&52.8&56.1&0.56&–&–&–&–&–\\
2-patch&83&20&18.3&9.7&48.5&0.86&45.0&56.1&0.22&–&–&–&–&–\\
2-patch&83&24&13.2&6.1&48.5&0.39&38.3&56.1&0.084&–&–&–&–&–\\
2-patch&83&28&9.5&3.8&48.5&0.18&32.7&56.1&0.033&–&–&–&–&–\\
2-patch&83&32&6.9&2.4&48.5&0.081&27.9&56.1&0.013&–&–&–&–&–\\
\hline
4-patch&31&12&70.0&67.1&75.5&35.4&75.9&81.6&22.0&23.0&25.3&1.54&0.96&0.87\\
4-patch&31&16&62.7&58.7&75.5&27.8&69.2&81.6&15.7&16.8&17.9&1.66&0.94&0.88\\
4-patch&31&20&56.2&51.4&75.5&21.8&63.2&81.6&11.2&12.7&12.6&1.71&0.88&0.89\\
4-patch&31&24&50.3&45.0&75.5&17.1&57.6&81.6&8.0&8.9&8.9&1.92&0.90&0.90\\
4-patch&31&28&45.1&39.4&75.5&13.4&52.6&81.6&5.7&5.6&6.3&2.37&1.02&0.92\\
4-patch&31&32&40.4&34.5&75.5&10.5&47.9&81.6&4.1&4.2&4.4&2.48&0.97&0.93\\
\hline
4-patch&83&12&34.9&27.7&48.5&4.7&58.8&56.1&1.5&2.4&2.7&1.98&0.65&0.57\\
4-patch&83&16&25.2&18.0&48.5&2.2&49.3&56.1&0.61&1.0&1.1&2.22&0.61&0.55\\
4-patch&83&20&18.3&11.8&48.5&1.0&41.3&56.1&0.24&0.41&0.46&2.56&0.59&0.52\\
4-patch&83&24&13.2&7.7&48.5&0.49&34.6&56.1&0.10&0.17&0.19&2.89&0.56&0.50\\
4-patch&83&28&9.5&5.0&48.5&0.23&29.0&56.1&0.038&0.070&0.079&3.30&0.54&0.47\\
4-patch&83&32&6.9&3.3&48.5&0.11&24.3&56.1&0.015&0.030&0.033&3.68&0.50&0.45\\
\hline\hline
\end{tabular}
\end{adjustbox}
\end{center}
\caption  {\small {Modeled and measured fidelities for the circuits of USTC's 31-qubit and 83-qubit random circuit sampling experiments conducted in 2024 \cite {Gao+25}.
(A) Single-qubit gate fidelity factor calculated from the individual (i.e. qubit-specific) gate errors in Figure~2a of Ref.~\cite {Gao+25}.
(B) Two-qubit gate fidelity factor calculated from the individual gate errors in Figure~2b of Ref.~\cite {Gao+25}.
(C) Readout fidelity factor calculated from the individual (i.e. qubit-specific) readout errors in Figure~2c of Ref.~\cite {Gao+25}.
(D) Modeled overall circuit fidelity according to Google's Formula (77) calculated as product of columns (A), (B), and (C).
(E) Idle gate fidelity for the 1-gates of those qubits that are not part of a two-qubit gate in two-qubit gate layers. Calculated from the average (not qubit-specific) values stated in Figure~S6b of the Supplement of Ref.~\cite {Gao+25}.
(F) Fidelity corresponding to a non-perfect preparation of the initial state $\lvert 000\ldots\rangle$ (called ``P0 error" in \cite {Gao+25}). Calculated from the average (not qubit-specific) values stated in Figure~S6c of the Supplement of Ref.~\cite {Gao+25}.
(G) Modeled overall circuit fidelity calculated as product of columns (D), (E), and (F).
(H) Measured XEB fidelity extracted from the data points in Figures~3 and S7 of Ref.~\cite {Gao+25} and its Supplement.
(I) Modeled ``estimated" fidelity extracted from the solid lines in Figures~3 and S7 of Ref.~\cite {Gao+25} and its Supplement.}}
\label{t:USTC_discrete_error_model}
\end{table}

\section {Proportion of measured 1's in NISQ samples}
\label {s:ratio}

\begin{figure*}[!ht]
    \centering
    \begin{minipage}{.5\textwidth}
        \centering
         \includegraphics[width=3.8in]{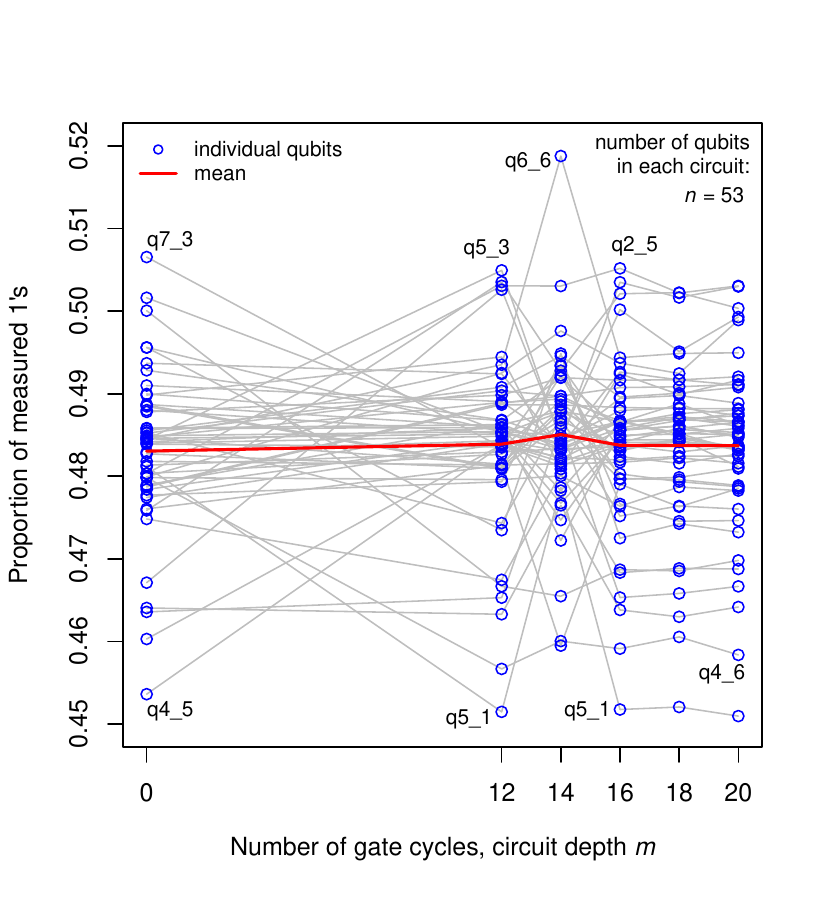}
    \end{minipage}%
    \begin{minipage}{.5\textwidth}
         \centering
         \hspace{1cm}
         \includegraphics[width=2.8in]{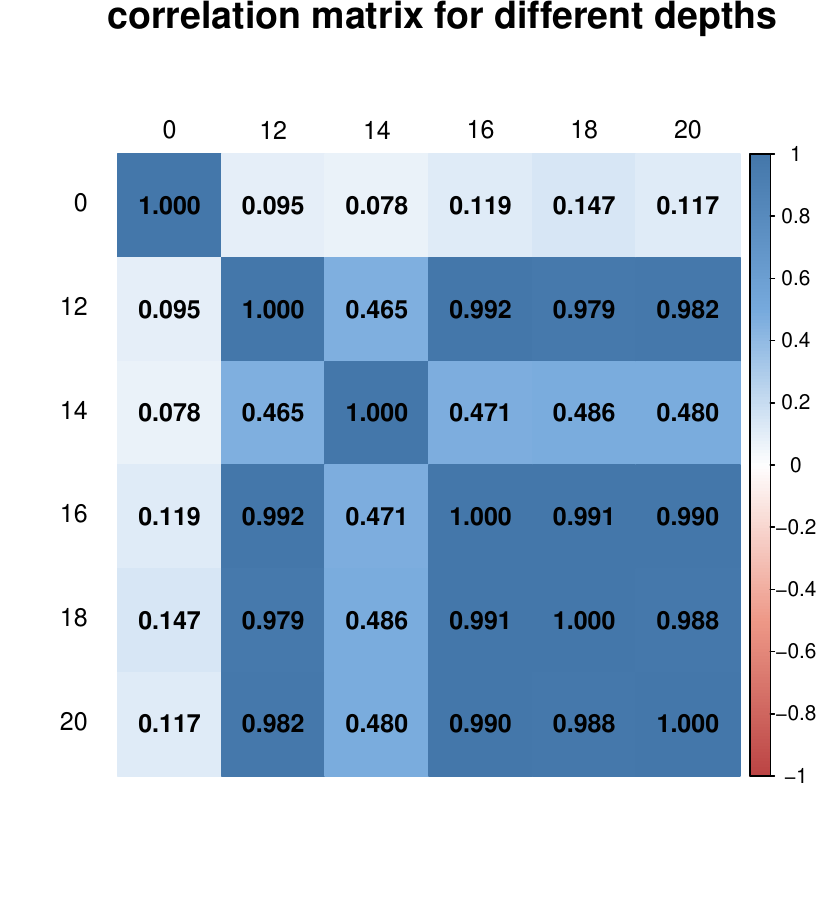}
    \end{minipage}
    \caption {\small {\textbf{Proportion of measured 1's as a function of the number of gate cycles (circuit depth $m$) for each of the Sycamore-53 individual qubits.}
    For depth $m=0$, the proportion of 1's was estimated from Google's initialization/measurement experiments (Google's published file \texttt{readout\_raw\_data.tar}) from the asymmetric readout error relative frequencies as $1/2 + (q_{0\rightarrow1}-q_{1\rightarrow0})/2$.
    For $m=0$, the positions of the individual qubits in the measured bitstrings are assumed to be the same in Google's initialization/measurement data \texttt{readout\_raw\_data.tar} as in the measured bitstrings of the full circuits (Python lists QUBIT\_ORDER in the published circuit description files).
    All published measurement bitstrings of Google's full circuits for number $n=53$ of qubits were considered, but not the elided and patch circuits.
    The slight increase of the proportion of measured 1's for many of the 53 qubits is contrary to an amplitude damping noise model, which predicts a decrease of this proportion with increasing number of gate cycles.
    A linear regression of the mean proportion results in a slope that is not significantly unequal to zero.
    This observation is in contrast to the more recent 67- to 70-qubit Sycamore and USTC's Zuchongzhi, for which this slope is significantly positive (see Figs.~\ref{fig_proportion_ones_SYC-70_vs_m} and \ref{fig_proportion_ones_ZCZ-56_vs_m}).
    The different behavior for depth $m=14$ reflects the different behavior between samples from ABCD circuits and EFGH circuits. (EFGH circuits contributed only for $m=14$).}}
    \label{fig_proportion_ones_SYC-53_vs_m}
\end{figure*}

In this appendix we study the proportion of measured 1's with increasing number of cycles in several NISQ experiment. Figure \ref{fig_proportion_ones_SYC-53_vs_m} describes the proportion of measured 1’s as a function of the number of gate cycles (circuit depth m) for each of the Sycamore-53 individual qubits, and compares this proportion to Google's data on readout errors from Appendix \ref {s:appA}. 

\begin{figure*}[!ht]
    \centering
    \hspace*{-1.2cm}
    \includegraphics[width=7.1in]{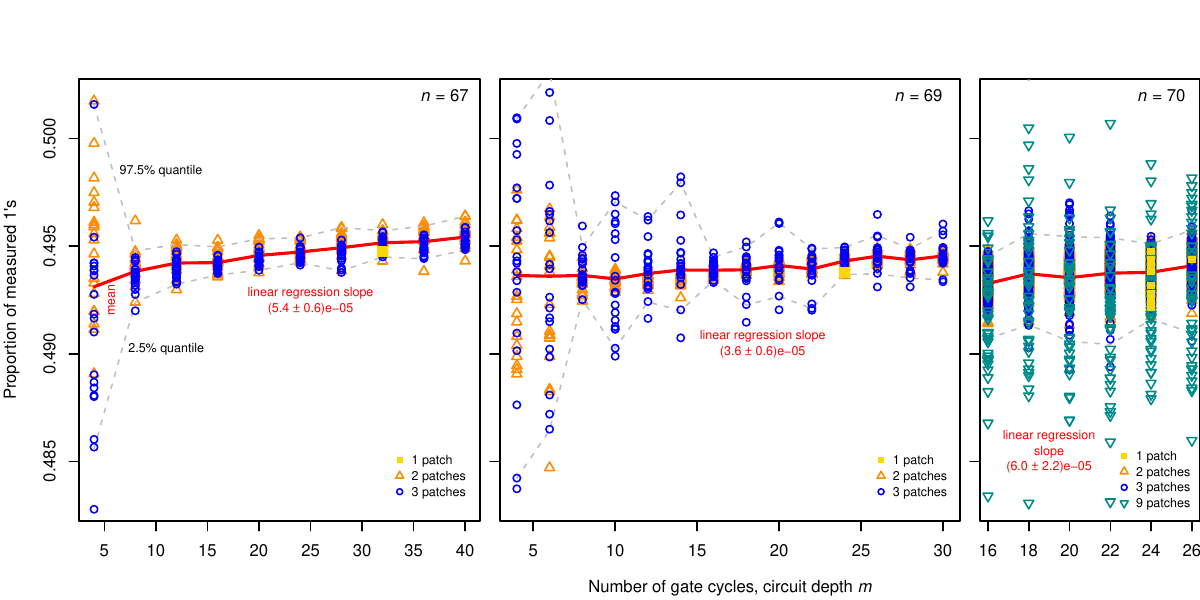}
    \caption{\textbf{Proportion of measured 1's as a function of the number of gate cycles (circuit depth $m$) for different circuits of Google's Sycamore processor with $n=67$ to $70$ qubits.}
    Each point corresponds to an individual circuit and shows the proportion value averaged over all bitstrings and all qubits for this circuit.
    The source of all data in this figure is Google's Zenodo data repository (version v3, 14 June 2024) to their 2023/24 publication ``Phase transitions in Random Circuit Sampling" (\cite {Mor+24}).
    For number of qubits $n=67$ and $n=69$ Google's data provide only one single full circuit (i.e. 1 patch, gold square) each.
    The linear regression slope $5.4 \times 10^{-5}$ for $n=67$ means that on average the proportion of 1's increases by this value for every additional gate cycle.
    This small but statistically significant increase of the mean proportion of measured 1's with increasing circuit depth $m$ is contrary to an amplitude damping noise model, which predicts a decrease of this proportion with increasing number of gate cycles.
    The standard deviation in the proportion of measured 1's is large for depth $m=4$ and becomes considerably smaller for larger $m$.}
    \label{fig_proportion_ones_SYC-70_vs_m}
\end{figure*}

\begin{figure*}[!ht]
    \centering
 \includegraphics[width=4.2in]{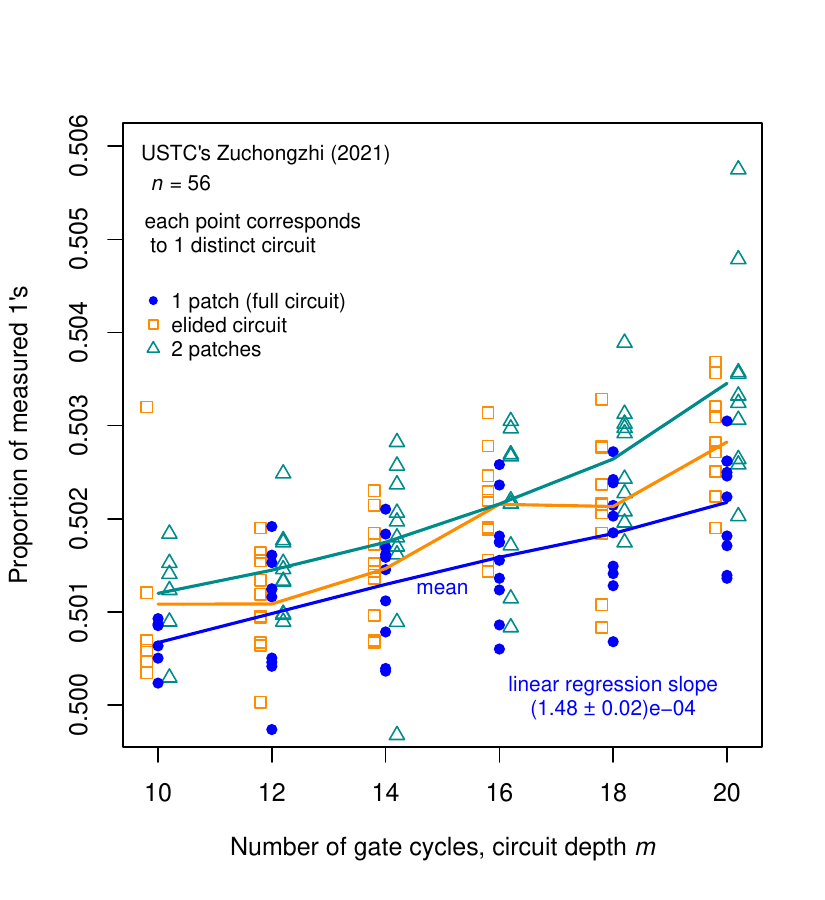}
    \caption{\textbf{Proportion of measured 1's as a function of the number of gate cycles (circuit depth $m$) for USTC's Zuchongzhi with 56 qubits.}
    Unlike Sycamore, Zuchongzhi's measured proportion of 1's are larger than 0.5.
    The mean proportion values for elided and patch circuits are larger than for full (i.e. 1-patch) circuits.
    The mean proportion of 1's for Zuchongzhi increases with increasing circuit depth $m$, which is similar to Sycamore's behavior for 67, 69, and 70 qubits.
    The slope of this increase for Zuchongzhi is significantly larger than for Sycamore (Figure~\ref{fig_proportion_ones_SYC-70_vs_m}).}
    \label{fig_proportion_ones_ZCZ-56_vs_m}
\end{figure*}

Figures \ref {fig_proportion_ones_SYC-70_vs_m} and \ref {fig_proportion_ones_ZCZ-56_vs_m} show an increase in the proportion of measured 1's, with an increasing number of cycles $m$ for Sycamore-67, -69, -70 and for USTC's Zuchongzhi-56.
This rise is contrary to findings by Fefferman et al. \cite{FGG+23} which predict a decrease in this proportion with increasing number of gate cycles.
In response to our inquiry on the matter, Yulin Wu from the USTC team attributed the proportion of 1's to leakage to non-computational basis states ($\lvert 2 \rangle$, $\lvert 3 \rangle$, etc.).

\comment {In response to our inquiry on the matter,
Yulin Wu from the USTC team wrote the following: 

``The reason why the proportion of 1’s is larger than 0.5 is due to leakage, or the `heating effect'.
In case of not perfectly calibrated gate leakage, adding more gate layers increases the probability of leaking to non-computational states progressively ($\lvert 2 \rangle$, $\lvert 3 \rangle$, etc.).

In sampling the random circuit, the readout is a two-state classification, these non-computational states are classified as $\lvert 0 \rangle$ or $\lvert 1 \rangle$ depends on the signal of leakage states are close to the signal of $\lvert 0 \rangle$ or $\lvert 1 \rangle$, in this particular experiment, they are classified as $\lvert 1 \rangle$, resulting in an increase of the proportion of 1’s. 

That is also why the proportion of 1’s is increasing progressively vs gate layers as shown in the your attached figure."
}

\clearpage

\section{Recent quantum computing experiments}
\label {s:newexp}
In Section C of the Appendix to \cite {KRS24} we listed some recent NISQ experiments that can be used for statistical analysis of samples from NISQ computers based on the tools of our papers or additional tools.
Here we update the list with a few additional papers.
The last two papers on the list do not refer to NISQ computers, but some ingredients of our statistical approach may be relevant also to them.   
\begin {enumerate}

\item A paper by the Google team  \cite {Google25a}  exhibits distance 3-, 5-, and 7- surface coded with better quality compared to their earlier works.

\item A new Google paper on the arXiv \cite {Google25b}
contains random circuit sampling data on a 103-qubit ``Willow" quantum processor.
The paper mentions the creation of samples that would require $10^{25}$ years on a classical supercomputer.
A related paper, also based on random circuits, \cite {Google25c} claims to achieve, for the first time, a verifiable quantum advantage.

\item Improved random sampling was also announced in several papers by researchers from USTC.
We already mentioned the recent paper by Gao et al. \cite {Gao+25}
describing a random sampling task performed with a 105-qubit Zuchongzhi 3.0 Processor, which would require $6.4 \times 10^9$ years to calculate on a classical supercomputer. 


\item We already mentioned Quantinuum's 2024 ``quantum supremacy" paper \cite {DHL+25} and their subsequent paper \cite {LSN+25} about certified randomness.
In Section C of the Appendix to \cite {KRS24}, we mentioned related results on ``quantum volume."
All of these experiments are characterized by very small samples.

\item A group from USTC \cite {Liu+25} announced a huge quantum advantage for Gaussian boson sampling with 3,050 photons and 8,176 modes.
The paper asserts the creation of samples (in only 25.6 microseconds by the Jiuzhang 4.0 photonic quantum computer) that would  require more than $10^{42}$ 
years on a classical supercomputer.

\item A new paper from Lidar's group \cite {Sin+25} claims that: ``Here, using two different 127-qubit IBM Quantum superconducting processors, we demonstrate an algorithmic quantum speedup for a variant of Simon’s problem." 

\item The Harvard/QuEra/MIT group has a new paper \cite {Blu+25} that utilizes reconfigurable arrays of up to 448 neutral atoms to implement key elements of a universal, fault-tolerant quantum processing architecture. 

\item A new paper by Kretschmer et al. \cite {Kre+25} 
describes a new form of quantum advantage, called ``quantum information supremacy," that represents a new benchmark in quantum computing and does not rely on conjectures from computational complexity.
The paper constructs a task, for which the most space-efficient classical algorithm provably requires between 62 and 382 bits of memory, and claims to solve it, on Quantinuum's H1-1 trapped-ion quantum computer, using only 12 qubits.

\item D-Wave's has a new paper \cite {Kin+25} claiming substantial quantum advantage.

\item Microsoft announced the first topological qubits based on Majorana zero modes.
This spans several papers including \cite {Microsoft25}.
\end {enumerate}

\section {Wish list} 
\label {s:wl}
We outline below several directions that, in our view, would facilitate a more accurate and comprehensive understanding of the current status of experimental quantum computing.

\subsubsection* {Samples from quantum computers} 
For our analysis, we require samples based on quantum circuits with a modest number of qubits, with particular emphasis on the 10–20 qubit range. It would be especially valuable to obtain samples for both random and other types of circuits in this range, using IBM’s quantum computers as well as other available platforms.

To date, we have not been able to obtain such samples from IBM researchers or colleagues with access to IBM’s quantum hardware, and we are making efforts to obtain such samples. It would also be important to assess the current experimental progress of quantum computers developed by other groups in both industry and academia and we have corresponded with a few such groups and approached others.

\subsubsection *{Additional simulations} 
Access to detailed simulated data under realistic noise models would substantially improve our ability to evaluate and interpret experimental claims.

\subsubsection* {Data sharing and independent verification} 
It would be desirable for researchers to make experimental data publicly accessible, ideally by depositing complete datasets on a public server at the time of publication. Furthermore, additional independent efforts are needed to critically examine published claims in quantum computing and to assess the corresponding data presented in support of those claims.

\end{document}